\documentclass[superscriptaddress,prx,twocolumn, longbibliography]{revtex4-2}
\usepackage{xcolor,amsthm,amsmath,amsxtra,amsfonts,dsfont,graphicx,bm}

\usepackage[utf8]{inputenc}
\usepackage[margin=2.5cm]{geometry}
\usepackage{amsmath, array}
\usepackage{amssymb}
\usepackage{physics}
\usepackage{graphicx}
\usepackage{qcircuit}
\usepackage[labelformat=simple]{subcaption}
\usepackage{setspace}
\usepackage{url,hyperref}
\hypersetup{
    colorlinks=true,
    linkcolor=blue,
    filecolor=blue,      
    urlcolor=blue,
    citecolor=blue
    }

\newcommand{\revise}[1]{\textcolor{black}{#1}}
\newcommand{\correct}[1]{\textcolor{black}{#1}}

\usepackage[english]{babel}
\addto\captionsenglish{} 
\addto\captionsenglish{} 
\usepackage[labelsep=period]{caption}

\usepackage[newcommands]{ragged2e}
\DeclareCaptionJustification{justified}{\justifying}
\DeclareUnicodeCharacter{2009}{-} 

\usepackage{chemformula} 

\begin{document}
\title{Qubit-efficient encoding scheme for quantum simulations of electronic structure}

\author{Yu Shee}
\affiliation{Department of Chemistry, University of California, Berkeley, California 94720, USA}

\author{Pei-Kai Tsai}
\affiliation{Department of Physics and Center for Theoretical Physics, National Taiwan University, Taipei 10617, Taiwan}

\author{Cheng-Lin Hong}
\affiliation{Department of Electrical Engineering and Graduate Institute of Communication Engineering, National Taiwan University, Taipei 10617, Taiwan}

\author{Hao-Chung Cheng}
\affiliation{Department of Electrical Engineering and Graduate Institute of Communication Engineering, National Taiwan University, Taipei 10617, Taiwan}
\affiliation{Department of Mathematics, National Taiwan University, Taipei 10617, Taiwan}
\affiliation{Quantum Computing Centre, Hon Hai (Foxconn) Research Institute, New Taipei City 236, Taiwan}
\affiliation{Center for Quantum Science and Engineering, National Taiwan University, Taipei 10617, Taiwan}
\affiliation{Physics Division, National Center for Theoretical Sciences, Taipei, 10617, Taiwan}

\author{Hsi-Sheng Goan} \email{goan@phys.ntu.edu.tw}
\affiliation{Department of Physics and Center for Theoretical Physics, National Taiwan University, Taipei 10617, Taiwan}
\affiliation{Center for Quantum Science and Engineering, National Taiwan University, Taipei 10617, Taiwan}
\affiliation{Physics Division, National Center for Theoretical Sciences, Taipei, 10617, Taiwan}

\begin{abstract}
  Simulating electronic structure on a quantum computer requires encoding of fermionic systems onto qubits. Common encoding methods transform a fermionic system of $N$ spin-orbitals into an $N$-qubit system, but many of the fermionic configurations do not respect the required conditions and symmetries of the system so the qubit Hilbert space in this case may have unphysical states and thus can not be fully utilized. We propose a generalized qubit-efficient encoding (QEE) scheme that requires the qubit number 
  to be only logarithmic in the number 
of configurations that satisfy the required conditions and symmetries.
For the case of considering only the particle-conserving and singlet configurations, we reduce the qubit count to \revise{an upper bound of} $\mathcal O(m\log_2N)$, where $m$ is the number of particles. This QEE scheme is demonstrated on
an $\ch{H2}$ molecule in
 the 6-31G basis set and a $\ch{LiH}$ molecule in the STO-3G basis set
using fewer qubits than the common encoding methods.
We calculate the ground-state energy surfaces using a variational quantum eigensolver algorithm  with a hardware-efficient ansatz circuit.
We choose to use a hardware-efficient ansatz since most of the Hilbert space in our scheme is spanned by desired configurations so a heuristic search for an eigenstate is sensible.
The simulations are
performed on \revise{IBM Quantum machines} and the Qiskit simulator with a noise model implemented from a IBM Quantum machine.
Using the methods of measurement error mitigation and error-free linear extrapolation, we demonstrate that most of the distributions of the extrapolated energies using our QEE scheme
agree with the exact results obtained by Hamiltonian diagonalization in the given basis sets within chemical accuracy.
Our proposed scheme and results show the feasibility of quantum simulations for larger molecular systems in the noisy intermediate-scale quantum (NISQ) era. \revise{The number of terms in Hamiltonian has an upper bound of $\mathcal O(\frac{N^{2m+1}}{(m-1)! \, m!})$ for the QEE scheme while it scales as $\mathcal O(N^{4})$ for the Jordan-Wigner encoding scheme. Nevertheless, we present several cases where QEE is useful.}
\end{abstract}

\date{\today}
\maketitle

\section{INTRODUCTION}
\label{sec: introduction}
The simulation of physical systems is one of the most prominent applications of quantum computing, as Feynman suggested to simulate a quantum system using another \cite{feynman_simulating_1982}. 
Much progress \cite{georgescu_quantum_2014, mcclean_theory_2016, mcardle_quantum_2020, abrams_quantum_1999} has been made since then, and several quantum algorithms for solving molecular energies have been proposed including quantum phase estimation (QPE) \cite{kitaev_quantum_1995, du_nmr_2010, lanyon_towards_2010, li_solving_2011, omalley_scalable_2016, paesani_experimental_2017, santagati_witnessing_2018, wang_quantum_2015, abrams_quantum_1999, aspuru-guzik_simulated_2005} and variational quantum algorithms (VQA) \cite{peruzzo_variational_2014, omalley_scalable_2016, mcclean_theory_2016, kandala_hardware-efficient_2017, kandala_error_2019, jiang_quantum_2018, kivlichan_quantum_2018, wecker_solving_2015, babbush_chemical_2015, sugisaki_quantum_2016, sugisaki_quantum_2019}.
The time cost of quantum simulation can scale polynomially with system size as compared to exponentially using classical computers \cite{mcardle_quantum_2020, aspuru-guzik_simulated_2005}.
With such polynomial time scaling of quantum algorithms and as quantum hardware is getting increasingly reliable and scalable, quantum simulation of physical systems has drawn great attention of researchers since classically intractable ab-initio calculations of proteins and materials could one day be realized using quantum bits (qubits) \cite{robert_resource-efficient_2021, mcclean_theory_2016, gao_computational_2021, cao_potential_2018, hempel_quantum_2018, colless_computation_2018, shen_quantum_2017}.

A significant challenge of quantum simulations is the quantum resources required to reliably perform quantum algorithms. Some previous works show that the number of qubits needed can be reduced \cite{moll_optimizing_2016, bravyi_tapering_2017} and how to execute quantum circuits on noisy intermediate-scale quantum (NISQ) devices \cite{nam_ground-state_2020}.
A practical approach for NISQ devices is the variational quantum eigensolver (VQE) algorithm \cite{peruzzo_variational_2014,mcclean_theory_2016,omalley_scalable_2016} which is a hybrid quantum-classical method to obtain the minimum eigenvalue of a given Hamiltonian operator. 
It requires a systematic encoding from fermionic systems to qubit systems and a preparation of trial states, or ansatzes, on the qubits.
For NISQ devices, the depth of an ansatz circuit should not be too deep
in case 
the quantum state 
operated by the circuit loses its coherence before the measurement process \cite{mcardle_quantum_2020, kandala_hardware-efficient_2017}, 
and a minimal number of two-qubit operations should be used 
as the errors in performing them are generally larger than those of single-qubit operations.

The encoding schemes also play important roles in the success of the VQE algorithm. 
Commonly used  encoding schemes such as the Jordan--Wigner (JW), parity, and Bravyi--Kitaev (BK) encoding methods often require $N$ qubits for a system with $N$ spin-orbitals \cite{mcardle_quantum_2020,jordan_1928, bravyi_fermionic_2002, seeley_bravyi-kitaev_2012}.
However, the number $N$ is often too large even for small molecules to be feasible on the present and near-term NISQ devices as 
the circuit depth and gate count of the ansatz circuit in the VQE algorithm usually also increases with $N$.  
Some compact encoding methods \cite{bravyi_tapering_2017, moll_optimizing_2016, babbush_exponentially_2018}
adopting symmetries or block diagonal features of Hamiltonian have also been proposed to reduce the number of qubits from $N$ to $\mathcal O(m\log_2N)$, where $m$ is the number of electrons.
Our work with the same reduction in the qubit number, described in detail later, is based on the second quantization formalism, whereas the works of Bravyi et al. \cite{bravyi_tapering_2017} and Babbush et al. \cite{babbush_exponentially_2018} are based on the first quantization method, and their encoding methods have a requirement on the number of particles, such that $m\log_2N < N$, for qubit reduction.
In Moll et al.'s work \cite{moll_optimizing_2016}, they proposed a scheme by first transforming the fermionic Hamiltonian into a qubit Hamiltonian by a common encoding scheme (e.g.~the JW or BK encoding scheme) and the reduced Hamiltonian is then obtained from a projected Hamiltonian using the block diagonal features of the fermionic Hamiltonian. However, their scheme needs to go through the whole qubit Hamiltonian using the reordering operators to reduce the number of qubits one by one, and the construction of the reordering operators is not generalized for systems with different dimensions and numbers of electrons. Therefore, these methods are not generalized as their implementations depend on the system size or particle number.


 Several methods \cite{kirby_second-quantized_2021, steudtner_fermion--qubit_2018} provide generalized schemes for qubit reduction where the implementations allow trade-off between qubit counts and gate counts. One of the methods presented in Steudtner and Wehner's work \cite{steudtner_fermion--qubit_2018} shows an exponential saving in qubit counts, but requires quantum devices with well-implemented multi-controlled gates. By considering the conservation of particles, Kirby et al. \cite{kirby_second-quantized_2021} presented an encoding method which reduces the length of BK encoded bitstrings with a given maximum allowed overlap of the codewords. The optimal encoding in the $N>\!\!>m$ limit has a polylogarithmic complexity in both qubit and gate counts \cite{kirby_second-quantized_2021}. The method from \cite{kirby_second-quantized_2021} has significant asymptotic savings but has not yet been demonstrated for quantum simulations, while in our work, we will be focusing on VQE algorithms for NISQ devices. 

Another method proposed by Di Matteo et al. \cite{di_matteo_improving_2021}
improves the Hamiltonian encoding with the Gray code. The work has a generalized encoding scheme but
considers only the two-body
problem of a deuteron, which can be reduced into an effective one-body problem involving only the relative motion of the neutron and proton of the deuteron. 
In other words, only the encoding of one-particle system is presented in \cite{di_matteo_improving_2021}.

In this paper, we propose a generalized qubit-efficient encoding (QEE) scheme to deal with many-electron (many-body) systems.
Instead of targeting the original system Hamiltonian, our QEE scheme aims at eliminating undesired electronic configurations, not only the configurations that do not respect the symmetries but also insignificant configurations found using perturbation arguments or other classical preprocessing methods from the system. 
This is similar to active space selection but the slight difference is that specific electronic configurations are chosen. 
Thus, the qubit counts can be further optimized as compared to previous commonly used methods, and the encoding of qubit Hamiltonian is at the last step of our scheme, so there is no need to obtain the original $N$-qubit Hamiltonian as in \cite{moll_optimizing_2016}. 
With the desired electronic configurations obtained, we then map these configurations to qubit basis states so we only need
the qubit number to be logarithmic in the number of
the desired configurations.
Finally, the qubit Hamiltonian can be constructed with the aid of operators that flip a qubit state or the operators that reflect the qubit state (entry-operators, as defined in Sec.~\ref{sec: Ham encoding}) which can be further decomposed to Pauli operator strings. Note that the antisymmetric fermionic exchange factors are directly taken into account from
the transition matrix elements of the excitation operators (including also the number operators) between the electronic configurations
(see Sec.~\ref{sec: Ham encoding} for details).



The rest of the paper is organized as follows. 
Sec.~\ref{subsec: QChem} briefly introduces the background of quantum computational chemistry. 
Sec.~\ref{subsec: encoding} formulates the QEE scheme we proposed. 
In Secs.~\ref{subsec: VQE} and \ref{subsec: error mitigation}, we demonstrated how to implement our method on quantum devices. Some examples for the QEE of molecules that do not use only a minimal basis set are shown in detail in Sec.~\ref{subsec: examples}. Sec.~\ref{sec: result} and Sec.~\ref{sec: conclusion} summarize and conclude our theoretical analysis and experimental results.

\section{METHOD}
\subsection{Quantum Chemistry and fermionic Hamiltonian}\label{subsec: QChem}
One of the most important problems in quantum computational chemistry is to find the
eigenvalues and eigenfunctions of the time-independent Schrodinger equation
\begin{equation}
    H\ket{\Psi}=E\ket{\Psi},
\end{equation}
where $\ket{\Psi}$ is an eigenfunction of the Hamiltonian operator $H$ with corresponding eigenvalues $E$. By applying the Born--Oppenheimer approximation, such a problem can be reduced to an electronic structure problem by treating the nuclei as fixed charges with electronic Hamiltonian being
\begin{equation}
     H=-\sum_i{\frac{\nabla_i^2}{2}}
    -\sum_{i, I}{\frac{Z_I}{\abs{r_i-R_I}}}
    +\frac{1}{2}\sum_{i\neq j}{\frac{1}{\abs{r_i-r_j}}},
\end{equation}
where $Z_I$ and $R_I$ denote the atomic number and position of the $I$th nucleus, and $r_i$ denotes the position of the $i$th electron. 
Alternatively, the second-quantization formalism of the electronic Hamiltonian projecting onto basis wavefunctions $\{\Psi_p(\mathbf{x_i})\}$ (with $\mathbf{x_i}$ being the spatial and spin coordinate of the $i$th electron) is
\begin{equation}
\label{eq:second_hamil_1}
  H_{\text{elec}}
  =\sum_{pq}
  {h_{pq}a_p^\dag a_q}
  +\frac{1}{2}\sum_{pqrs}
  {h_{pqrs}a_p^\dag a_q^\dag a_ra_s},
\end{equation}
where $h_{pq}$ and $h_{pqrs}$ are the one- and two-electron integrals defined as
\begin{equation}
h_{pq} = \int d\mathbf{x}\Psi^*_p(\mathbf{x}) \left( -\frac{\nabla^2}{2}
    -\sum_{I}{\frac{Z_I}{\abs{r-R_I}}} \right) \Psi_q(\mathbf{x}),
\end{equation}
\begin{equation}
h_{pqrs} = \int d\mathbf{x}_1 d\mathbf{x}_2 \frac{\Psi^*_p(\mathbf{x}_1)\Psi^*_q(\mathbf{x}_2)\Psi_r(\mathbf{x}_2)\Psi_s(\mathbf{x}_1)}{\abs{\mathbf{x}_1-\mathbf{x}_2}}.
\end{equation}
The creation and annihilation operators are defined as
\begin{align}
\begin{split}
  a_p^\dag 
  & \ket{f_{N-1}, \cdots, f_p,\cdots, f_0} \\
  & = \delta_{f_p, 0}
  (-1)^{\sum_{i=0}^{p-1} f_i}
  \ket{f_{N-1}, \cdots, 1\oplus f_p,\cdots, f_0}
\end{split}
\end{align}
\begin{align}
\begin{split}
  a_p 
  & \ket{f_{N-1}, \cdots, f_p,\cdots, f_0} \\
  & = \delta_{f_p, 1}
  (-1)^{\sum_{i=0}^{p-1} f_i}
  \ket{f_{N-1}, \cdots, 1\oplus f_p,\cdots, f_0},
\end{split}
\end{align}
where $\ket{f_{N-1}, \cdots, f_p,\cdots, f_0}$ is a vector of occupation numbers representing whether an electron is presenting ($f_p = 1$) in the Slater determinant or not ($f_p = 0$), the $\oplus$ sign denotes addition modulo 2, and the term $(-1)^{\sum_{i=0}^{p-1} f_i}$ addresses the exchange antisymmetric nature of fermions \cite{mcardle_quantum_2020, szabo_modern_1996}.

\subsection{Encoding}\label{subsec: encoding}
To simulate fermionic systems on quantum processors, an encoding of fermionic states is needed. 
Following \cite{bravyi_tapering_2017}, an encoding is an isometry $\mathcal E: \mathcal H_{\text{elec}} \to \mathcal H_{\text{q}}$ where $\mathcal H_{\text{elec}}$ and $\mathcal H_{\text{q}}$ denote fermionic and qubit Hilbert space. 
Note that in this work, a ket state with subscript f or q indicates that it is a fermionic or qubit state, respectively.
A fermionic state $\ket{\mathbf f}_{\text{f}} = \ket{f_{N-1},...,f_0}_{\text{f}} \in \mathcal H_{\text{elec}}$ corresponds to a qubit state $\mathcal E \ket{\mathbf f}_{\text{f}} \in \mathcal H_{\text{q}}$, and a fermionic Hamiltonian $H_{\text{elec}}$ is mapped to its qubit counterpart $H_{\text{q}}\equiv \mathcal E \circ H_{\text{elec}} \circ \mathcal E^{-1}$.

For example, the mapping of the JW encoding method is defined as $\mathcal E_\text{JW} \ket{\mathbf f}_{\text{f}} \equiv \ket{\mathbf f}_{\text{q}}$. That is, the $i$th qubit represents the occupation number of the $i$th spin-orbital. The mapping of creation and annihilation operators under JW transform are
\begin{align}
    a_p 
    & \to Q_p \otimes Z_{p-1}\otimes\cdots\otimes Z_0, \\
    a_p^\dag 
    & \to Q_p^\dag \otimes Z_{p-1}\otimes\cdots\otimes Z_0,
\end{align}
where $Q_p^\dag=\frac12 (X_p-iY_p)$ and $Q_p = \frac12 (X_p+iY_p)$ are qubit creation and annihilation operators, respectively. For the rest of the work, we often mention common encoding schemes including the JW, parity, and the BK fermionic-to-qubit mappings.

\subsubsection{Qubit-Efficient Encoding}
Common fermionic-to-qubit mappings often describe a qubit Hamiltonian with a spin-orbital basis where some segments of the Hilbert space are spanned by insignificant electronic configurations such as configurations with incorrect particle numbers. As Eq.~(\ref{eq:second_hamil_1}) suggests that the second-quantized Hamiltonian is particle-conserving, configurations with incorrect particle numbers should not contribute to energy expectation values (Eq.~(\ref{eqn:VariationalPrinciple})). 
Using a basis with relevant configurations to describe qubit Hamiltonians will reduce the number of qubits required for simulating fermionic many-body systems. 
We show how to systematically construct qubit Hamiltonians from the configurations that respect total particle numbers. We also show that the total spin of the configurations can be restricted to further reduce the number of qubits required.

For a system with $N$ spin-orbitals and $m$ electrons, there are only $N\choose m$ particle-conserving electronic configurations. Instead of using $N$ qubits, we expect that $Q=\left\lceil\log_2{N\choose m}\right\rceil$ qubits is enough for simulation. 
First, we define a fermionic configuration $\ket {\mathbf f}_{\text{f}} = \ket{f_{N-1},...,f_0}_{\text{f}}$ and the set of all particle-conserving fermionic configurations
$\mathcal F_m
= \{ \ket {\mathbf f} _{\text{f}} \in\mathcal H_{\text{elec}}:|\mathbf f| = m \}$ 
where $|\mathbf f| = |\{k: f_k=1\}|$ denotes the Hamming weight, i.e., the total number of 1 (non-zero element), of $\mathbf f$.
Configurations in $\mathcal F_m$ are mapped to 
$\mathcal Q_Q=\{\ket{0}_{\text{q}}, \ket{1}_{\text{q}}\}^{\otimes Q}$,
the computational basis states of a $Q$-qubit system, by $\mathcal E$. On the other hand, similar to the concepts of freezing or removing insignificant orbitals, we can remove or add any electronic configurations into the $\mathcal F_m$ set so that the $Q$-qubit Hilbert space is optimally exploited. For example, even though we have chosen all the particle-conserving configurations to be in $\mathcal F_m$, some of the configurations might not be contributing to electronic correlation by using perturbation theory arguments. Thus, these configurations can be removed from the set and we can still map the rest of the configurations ascendingly to a $Q$-qubit Hilbert space or even smaller Hilbert space.

The choice of $\mathcal E$ can be various as long as it is an isometry. Previous works have used Gray-code encoding for $m=1$ system \cite{di_matteo_improving_2021} or sparse encoding to construct sparse qubit Hamiltonian \cite{bravyi_tapering_2017}.
As a configuration $\ket {\mathbf f}_{\text{f}} = \ket{f_{N-1},...,f_0}_{\text{f}}$ can be represented by a decimal number \revise{No.} $ \ket {\mathbf f}_{\text{f}} \equiv \sum_{i=0}^{N-1} f_i 2^i$, 
we sort $\mathcal F_m$ in an ascending order such that $\mathcal F_m = \{\ket{\mathbf f_0}_{\text{f}}, \ket{\mathbf f_1}_{\text{f}},...\}$ 
with \revise{No.} $\ket{\mathbf f_0}_{\text{f}}< \revise{\text{No. }}\ket{\mathbf f_1}_{\text{f}}<\cdots$. 
Similarly, states in $\mathcal Q_Q$ can be also sorted in an ascending order as $\mathcal Q_Q = \{\ket{\mathbf q_0}_{\text{q}}, \ket{\mathbf q_1}_{\text{q}}, ...\}$.
In this work, we define QEE as $\mathcal E_{\text{QEE}}\ket{\mathbf f_i}_{\text{f}} = \ket{\mathbf q_i}_{\text{q}}$. 
Even though the mapping of fermioinic configurations to qubit basis states could be in an arbitrary order, we align $\mathcal F_m$ with $\mathcal Q_Q$ in an ascending manner (from Hartree--Fock configuration to the fully-excited configuration). 
In this way, the asymmetric state preparation and measurement (SPAM) error of the $\ket{0}$ and $\ket{1}$ qubit states can be taken into account.
Since the $\ket{1}$ state is often more error-prone than the $\ket{0}$ state in real quantum devices (see, e.g.~, the last two columns of Table \ref{tab:santiago_single} in Appendix \ref{appendix:calibration}), we choose an ascending encoding such that more significant configurations are represented by qubit states with more bits in the $\ket{0}$ state to increase the fidelity of our computation. 
Note that this encoding does not require single-qubit gates for initialization as in other encoding methods because the $\ket{0}^{\otimes Q}$ state is the Hartree--Fock state that can be our reference state for post-Hartree--Fock methods.

\subsubsection{Hamiltonian Encoding}\label{sec: Ham encoding}
In most common fermionic-to-qubit mapping schemes, both creation and annihilation operators have their corresponding qubit operators. 
However, a single creation or annihilation operator changes the number of electrons in the system, which results in irrelevant electronic states and cannot be encoded using QEE. 
We first rewrite the second-quantized Hamiltonian as 
\begin{align}
    & H_{\text{elec}}
    = \sum_{pq}
    {h_{pq}a_p^\dag a_q}
    +\frac{1}{2}\sum_{pqrs}
    {h_{pqrs}a_p^\dag a_q^\dag a_ra_s}
    \label{eq:second_hamil}
    \\
    & = \sum_{pq}{h_{pq}E_{pq}}
    +\frac12 \sum_{pqrs} h_{pqrs}(
\correct{     \delta_{qr}E_{ps}  } 
    -E_{pr}E_{qs})
    \label{eq:double_excitation_identity}
\end{align}
where we define excitation operators $E_{pq}\equiv a_p^\dag a_q$ (which also include number operators when $p=q$) and use the fermionic anti-commutation relation $\{a_p^\dag, a_q\} = \delta _{pq}$.
As $E_{pq}$ is a particle-number-conserving operator, the electronic Hamiltonian expressed in terms of excitation operators can be mapped to qubit operators with QEE.

Any excitation operator $E_{pq}$ can be written as $E_{pq} = \sum_{k,k'=0}^{|\mathcal F _m|-1} 
c_{k'k}^{pq} \ket{\mathbf f_{k'}}_{\text{f}} \bra{\mathbf f_k}_{\text{f}}$, where
$c^{pq}_{k'k} = \bra{\mathbf f_{k'}}_{\text{f}} E_{pq} \ket{\mathbf f_k}_{\text{f}}$ is the corresponding coefficient.
$c^{pq}_{k'k}$ is zero if the transition from $\ket{\mathbf f_k}_{\text{f}}$ to $\ket{\mathbf f_{k'}}_{\text{f}}$ via $E_{pq}$ is impossible; 
otherwise it can be $\pm 1$ due to the antisymmetric nature of fermions. 
In particular, for 
$\ket{\mathbf f_{k}}_{\text{f}} = \ket{f_{N-1},..., f_p=0,..., f_q=1, ..., f_0}$
and 
$\ket{\mathbf f_{k'}}_{\text{f}} = \ket{f_{N-1},..., f_p=1,..., f_q=0, ..., f_0}$,
the coefficient $c^{pq}_{k'k}$ is plus/minus one if the sum of $f_i$ in between $f_p$ and $f_q$ is even/odd,
i.e., $c^{pq}_{k'k} = \prod_{i=\min(p,q)+1}^{\max(p,q)-1} (-1)^{f_i}$.
Note that
we decompose $E_{pq}$ into a linear combination of transitions from $k$ to $k'$,
but for each $k$, only one $k'$ gives non-zero $c^{pq}_{k'k}$.

As any excitation operator $E_{pq}$ can be decomposed as $E_{pq} = 
\sum_{k,k'=0}^{|\mathcal F _m|-1} 
c_{k'k}^{pq} \ket{\mathbf f_{k'}}_{\text{f}} \bra{\mathbf f_k}_{\text{f}}$,
the corresponding qubit operator is 
$\Tilde{E}_{pq} = \mathcal E \circ E_{pq} \circ \mathcal E^{-1}
= \sum_{k, k'=0}^{|\mathcal F _m|-1} 
c^{pq}_{k'k} \ket{\mathbf q_{k'}}_{\text{q}}\bra{\mathbf q_k}_{\text{q}}$,
where $\ket{\mathbf q_k}_{\text{q}} = \mathcal E \ket{\mathbf f_k}_{\text{f}}$ is the encoded qubit state of $\ket{\mathbf f_k}_{\text{f}}$.
In our encoding scheme, all configurations in $\mathcal F_m$ are mapped to some $Q$-qubit computational basis states. 
Hence for two computational basis states $\ket{\mathbf q}_{\text{q}} = \ket{q_{Q-1},...,q_0}_{\text{q}}$ 
and $\ket{\mathbf q'}_{\text{q}} = \ket{q'_{Q-1},...,q_0'}_{\text{q}}$,
the transition $\ket{\mathbf q'}_{\text{q}}\bra{\mathbf q}_{\text{q}}$ can be factorized as ${\bigotimes_{k=0}^{Q-1}} \ket{q'_k}_{\text{q}} \bra{q_k}_{\text{q}}$. We further define the qubit creation operator $Q^+$, qubit annihilation operator $Q^-$, qubit number operators $N^{(0)}$ and $N^{(1)}$ as
\begin{align}
    Q^+ & = \ket{1}_{\text{q}} \bra{0}_{\text{q}} = \frac12 (X-iY), \\
    Q^- & = \ket{0}_{\text{q}} \bra{1}_{\text{q}} = \frac12 (X+iY), \\
    N^{(0)} & = \ket{0}_{\text{q}} \bra{0}_{\text{q}} = \frac12 (I+Z), \\
    N^{(1)} & = \ket{1}_{\text{q}} \bra{1}_{\text{q}} = \frac12 (I-Z). 
\end{align}

In this work, these four operators are called entry-operators as each of them has exactly one non-zero entry in its matrix representation. 
A qubit state transition can thus be written as a tensor product of some entry-operators.
Then any (encoded) excitation operator $\Tilde{E}_{pq}$ can be expressed as a sum of the products of entry-operators,
\begin{equation}\label{eq: Epq as T}
    \Tilde{E}_{pq} 
    = \sum_{k,k'=0}^{|\mathcal F _m|-1} 
    c^{pq}_{k'k} 
    \ket{\mathbf q_{k'}}_{\text{q}}
    \bra{\mathbf q_k}_{\text{q}}
    = \sum_{k,k'=0}^{|\mathcal F _m|-1}
    \bigotimes_{w=0}^{Q-1} c^{pq}_{k'k}  T_{k'k,w},
\end{equation}
where $T_{k'k,w}$ are some entry-operators corresponding to $\ket{\mathbf q_{k'}}_{\text{q}}\bra{\mathbf q_k}_{\text{q}}$.
As each entry-operator is a sum of two Pauli operators, Eq. (\ref{eq: Epq as T})~can be then expressed in terms of Pauli operator strings, which allows us to write down the qubit Hamiltonian
\begin{equation}
{H}_{\text{q}} = \sum_{pq}{h_{pq}\Tilde E_{pq}}
    +\frac12 \sum_{pqrs} h_{pqrs}(
\correct{ \delta_{qr}\Tilde E_{ps} }
    -\Tilde E_{pr} \Tilde E_{qs})
\end{equation}
as sum of Pauli operator strings. 
Finally, the expectation value of ${H}_{\text{q}}$ can be evaluated and minimized on a quantum processor with variational quantum algorithms.



\subsection{Variational Quantum Eigensolver and Ansatz Circuit}
\label{subsec: VQE}


After obtaining the qubit Hamiltonian, we then 
apply the
VQE algorithm to solve the electronic structure problem on a quantum processor. Given a qubit Hamiltonian $H$ with unknown minimum eigenvalue $E_{\text{min}}$ and its corresponding eigenstate $\left | \Psi_{\text{min}} \right\rangle$, the variational method in Eq. (\ref{eqn:VariationalPrinciple})~helps find the ground-state energy by tuning $\bm{\theta}$ (note that $\bm{\theta}$ represents a vector that contains one or more parameters) in the parametrized trial wavefunction $ \left | \Psi(\bm{\theta}) \right\rangle$. 
\begin{equation}
\label{eqn:VariationalPrinciple}
    E(\bm{\theta}) 
    \equiv 
    \left\langle \Psi(\bm{\theta}) \right |
     H 
    \left | \Psi(\bm{\theta}) \right\rangle
    \ge E_{\text{min}}.
\end{equation}
In other words, the ground-state energy and wavefunction can be found by finding the parameters that minimize the energy expectation value $\left\langle \Psi(\bm{\theta}) \right | H \left | \Psi(\bm{\theta}) \right\rangle$. Such a VQE algorithm is separated into hybrid executions for quantum and classical computers, where the trial wavefunctions preparation and measurement are done on a quantum processor, and the parameters are updated classically using optimization algorithms.

The parametrized wavefunction can be represented as
\begin{equation}
\label{eqn:AnsatzReference}
  \ket{\Psi(\bm{\theta})} 
  = U(\bm{\theta}) \ket{\Psi_{\text{ref}}},
\end{equation}
where $\ket{\Psi_{\text{ref}}}$ is a reference state that has great overlap with the minimum eigenstate $\left | \Psi_{\text{min}} \right\rangle$ and $U(\bm{\theta})$ is the trial state circuit (or ansatz circuit). 
Qubit registers are often initialized into the $\ket{0}^{\otimes Q}$ state and we often need to apply unparametrized operations on the $\ket{0}^{\otimes Q}$ state to obtain a reference state $\ket{\Psi_{\text{ref}}}$ that properly describes the chemical systems. For example, the reference state could be a Hartree--Fock state or a multi-reference state \cite{jiang_quantum_2018, kivlichan_quantum_2018, wecker_solving_2015, babbush_chemical_2015, sugisaki_quantum_2016, sugisaki_quantum_2019}. Since the $\ket{0}^{\otimes Q}$ state is the Hartree--Fock state in our method so we do not need to apply any unparametrized operations. For the measurement process of the VQE algorithm, since a qubit Hamiltonian is a
linear combination of Pauli strings, the linearity of expectation can thus be used to calculate
the energy expectation value
\begin{equation}
    \left\langle \Psi(\bm{\theta}) \right 
    | H 
    \left | \Psi(\bm{\theta}) \right\rangle 
    = \sum_i h_i \left\langle \Psi(\bm{\theta}) \right | P_i \left | \Psi(\bm{\theta}) \right\rangle
\end{equation}
for each trial state with a given set of parameters $\bm{\theta}$. 

The design of circuit ansatz determines the feasibility and accuracy of the VQE algorithm. 
To capture or approximate the complexity of the exact wavefunction, researchers came up with chemically-inspired quantum circuit design with inspiration from classical computational chemistry. 
The most common one is the unitary coupled cluster (UCC) ansatz stemming from classical coupled cluster theory \cite{bartlett_alternative_1989, hoffmann_unitary_1988}. 
However, chemically-inspired approaches often need Trotter approximation, and the depth of the circuits is often too large to be practical on NISQ devices.
Another common type of circuit ansatz design is the hardware-efficient approach where it aims at heuristically describing a wavefunction with
thorough
consideration of hardware constraints for NISQ devices \cite{peruzzo_variational_2014, kandala_hardware-efficient_2017, kandala_error_2019}. 
Hardware-efficient circuits are often built up by single-qubit rotation operators and few numbers of nearest neighbor CNOT gates. 
This type of circuit ansatz can also capture the complexity of the exact wavefunction because of its high expressibility and entangling capability where the parameterized circuits span most of the Hilbert space including the chemical subspace required to find minimum eigenenergies \cite{sim_expressibility_2019}. 
The concern of using hardware-efficient ansatz for qubit Hamiltonian obtained from common mapping schemes is that the parameters often get stuck on barren plateaus or local minimum where gradients vanish in this region during optimization. 
This is because these heuristic ansatz circuits are often used to solve problems
by trial and error method with repeated, varied attempts.
In the case of using VQE to solve electronic structure problem, the span of the hardware-efficient parameterized circuits often includes not only appropriate symmetry subspace but also insignificant sectors of the Hilbert space.  Thus, the final state obtained from hardware-efficient ansatz using common mapping schemes might include configurations that do not respect the conservation of particle numbers or other symmetries. 

Using QEE, most of the Hilbert space is spanned by significant configurations with appropriate symmetry. If the number of fermionic configurations that respect the required symmetry is $|\mathcal F _m| = 2^Q$, i.e., we can use exactly $Q$ qubits to encode all the configurations
so no insignificant configurations will be in
$Q$-qubit Hilbert space
$\mathcal H_{\text{q}}$ (see, e.g.~, Table \ref{tab:H2_STO3G_restricted_mapping}). Whereas if $2^{Q-1} < |\mathcal F _m| < 2^Q$, since we still need a $Q$-qubit Hilbert space to accommodate all the fermionic configurations, there will be some qubit basis states not representing any of the desired fermionic configurations (see, e.g.~, Table \ref{tab:H2_STO3G_unrestricted_mapping}). Nevertheless, the number of insignificant configurations presenting in such a $Q$-qubit $\mathcal H_{\text{q}}$ is still much fewer than that of in a $N$-qubit Hilbert space using common mapping or encoding schemes for the case of $2^{Q-1} < |\mathcal F _m| < 2^Q$.
Thus, we can use hardware-efficient ansatz circuits since the final states are more likely to respect the required symmetry. 
Also, it is stated in \cite{mcclean_barren_2018} that the gradients of parameterized circuits vanish exponentially as qubit counts increase.
Therefore, our encoding method would suffer less from the barren plateau problem comparing to common encoding schemes that require exponentially more qubits than our QEE method. 
More importantly, our encoding method reduces the coherence time required for operating the quantum circuit in the VQE algorithm as compared to common mapping schemes that require a larger number of qubits and consequently a higher ansatz circuit depth to reach the same level of entanglement. 

\begin{figure}[hbt!]
\centerline{  
\includegraphics[width=0.5\textwidth]{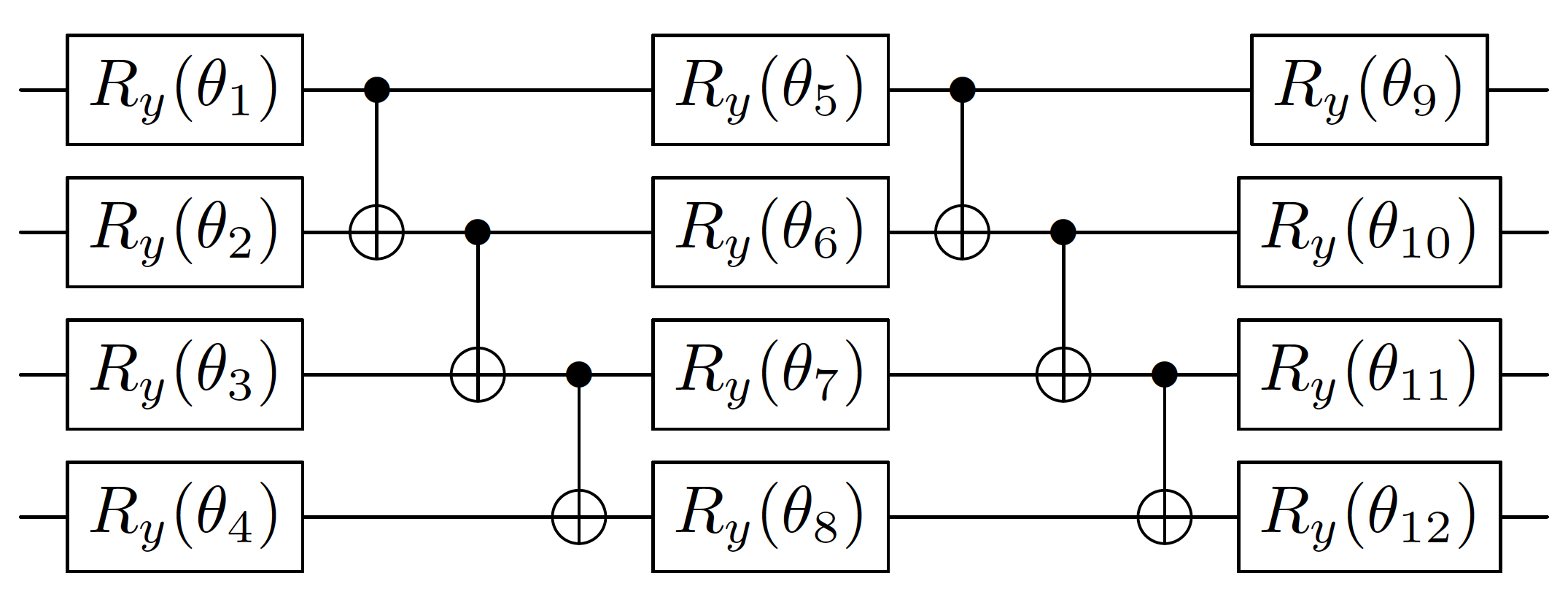}
}
\caption{4-Qubit ansatz circuit without redundant CNOT gates.}
\label{fig:AnsatzCircuit6}
\end{figure}

For the illustrative example in Sec.~\ref{sec: H2_631G_illustrative} and Sec.~\ref{sec: LiH_STO3G_illustrative}, we use a 4-qubit hardware-efficient ansatz circuit that consists of alternating layers of $R_y$ rotations and CNOT entanglements shown in Fig.~\ref{fig:AnsatzCircuit6}. This is because the circuit depth for UCC scales at least quadratically with respect to the number of qubits \cite{barkoutsos_quantum_2018} compared to linear scaling for our hardware-efficient ansatz, so it might not be practical to use UCC ansatz circuits in this case.
The number of repetitions of the alternating layers is two with an additional final rotation layer, and the entanglement pattern uses only the nearest neighbor CNOT gates. Since only  $R_y$  and CNOT gates are employed in the circuit, the prepared quantum states will only have real amplitudes. This circuit is also called the real-amplitudes 2-local circuit \cite{noauthor_qiskitcircuitlibraryrealamplitudes_nodate}.

\subsection{Error Mitigation}
\label{subsec: error mitigation}

\begin{figure*}[t]
\centerline{
\includegraphics[width=0.8\textwidth]{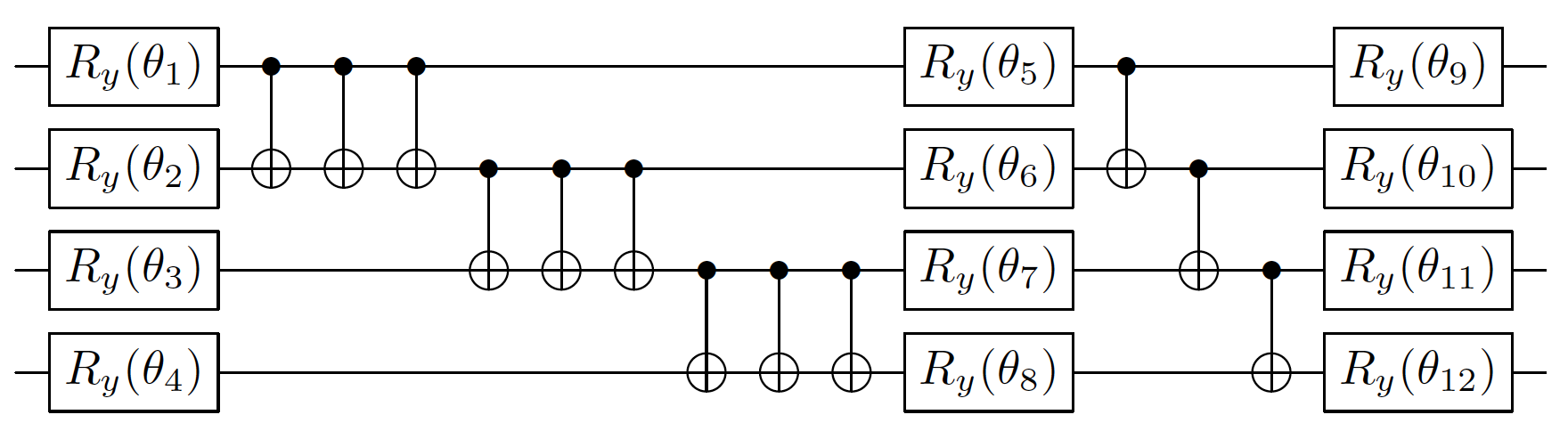}
}
\caption{4-Qubit ansatz circuit with 6 redundant CNOT gates.}
\label{fig:AnsatzCircuit12}
\end{figure*}

\begin{figure*}[t]
\centerline{
\includegraphics[width=\textwidth]{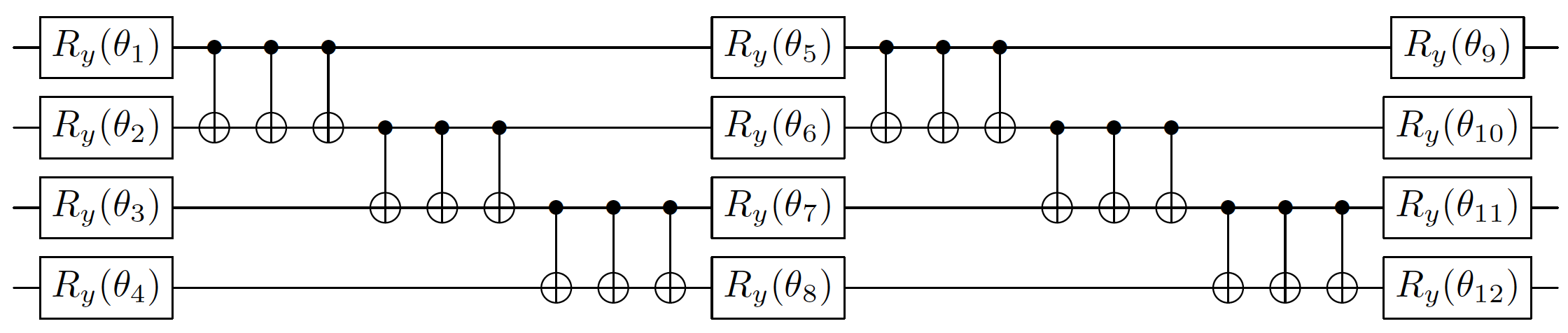}
}
\caption{4-Qubit ansatz circuit with 12 redundant CNOT gates.}
\label{fig:AnsatzCircuit18}
\end{figure*}

Errors in near term quantum processors accumulate quickly during computations which could ruin the results or, specifically, energy expectation values for the VQE algorithm. 
Quantum error correction methods could help fix this problem, but these methods often require a large number of qubits that would be impractical on near-term NISQ devices. 
Even though compared to common mapping schemes, our encoding method has reduced the number of qubits and lowered the circuit depth, the error in noisy hardware is still considerable, preventing us from reaching the desired accuracy of the observables.
Therefore, we adopt error mitigating methods in our illustrative examples to demonstrate the feasibility of our method. 

The first error mitigating method we use is measurement error mitigation. It calibrates readout counts by applying the inverse of the matrices generated from measurement calibration circuits from each basis state \cite{temme_error_2017, endo_practical_2018}.
The other method we adopt is an extrapolation method. This is done by amplifying the major error rate of the ansatz circuit and approximate the error-free limit of the energy expectation value by linear extrapolation \cite{kandala_error_2019, li_efficient_2017, temme_error_2017, endo_practical_2018}.
For example, the noise of the ansatz circuit in Fig.~ \ref{fig:AnsatzCircuit6} is dominated by two-qubit CNOT gate errors so we can perform the VQE algorithm using similar constructs of the ansatz circuit but with some redundant CNOT pairs as shown in Fig.~ \ref{fig:AnsatzCircuit12} and Fig.~ \ref{fig:AnsatzCircuit18}. 
We can thus assume that the observable depends on a noise variable $\epsilon=c\epsilon_0$, where $c$ is the number of CNOT gates, and $\epsilon_0$ is the error of a single CNOT gate (assuming the same CNOT gate error for every qubit). 
After extrapolating the results from each circuit, we could get the observable by linear extrapolation method at error-free (no CNOT gate error) condition. This is done by assuming that the CNOT gate error is still small enough so that we could do a first-order Taylor expansion of the noisy observable $O(\epsilon)$ with respect to $\epsilon$ at $\epsilon \approx 0$,
\begin{equation}
    O(\epsilon)\approx O(0) + c\epsilon_0 O'(0),
\end{equation}
so that we would see a linear response of the observable error to the CNOT gate error and can find the error-free observable $O(0)$. 

\subsection{Illustrative Examples}
\label{subsec: examples}
\subsubsection{$\ch{H2}$, STO-3G, Total-spin-restricted}
\label{subsubsec: H2_total_spin_restricted}
We show here a step-by-step instruction of how QEE maps the excitation operators $E_{pq}$ to qubit operators in Pauli operator strings.
We use an $\ch{H2}$ molecule in the STO-3G basis set as an example. 
In this case, the system includes only two $1s$ atomic orbitals so there are 4 converged spin-orbitals ($\sigma_{1s, u\downarrow}, \sigma_{1s, g\downarrow}, \sigma_{1s, u\uparrow}, \sigma_{1s, g\uparrow}$) from a Hartree--Fock self-consistent field calculation. 
We can write an electronic configuration in the fermionic occupation basis of this system as
\begin{equation}
    \ket{f_{\sigma_{1s, u\downarrow}}, f_{\sigma_{1s, g\downarrow}}, f_{\sigma_{1s, u\uparrow}} f_{\sigma_{1s, g\uparrow}}}
    = \ket{f_3, f_2, f_1, f_0}.
    \label{f_restricted}
\end{equation}
Furthermore, we restrict the total spin
such that only the singlet electronic configurations are present in the system. 
We can then map these fermionic configurations in an ascending order into the qubit state basis as shown in Table \ref{tab:H2_STO3G_restricted_mapping}.

\begin{table}[hbt!]
\centering
\caption{Mapping of fermionic configurations to qubit basis state for a total-spin-restricted $\ch{H2}$ molecule in the STO-3G basis set.}

\begin{ruledtabular}
\begin{tabular}{ccc}
Filled spin-orbitals&$f_3f_2\,f_1f_0$&$q_1\,q_0$\\
\hline
$\sigma_{1s, g\downarrow}\sigma_{1s, g\uparrow}$ & 01\,\,01 & 0\,\,0 \\
$\sigma_{1s, g\downarrow}\sigma_{1s, u\uparrow}$ & 01\,\,10 & 0\,\,1 \\
$\sigma_{1s, u\downarrow}\sigma_{1s, g\uparrow}$ & 10\,\,01 & 1\,\,0 \\
$\sigma_{1s, u\downarrow}\sigma_{1s, u\uparrow}$ & 10\,\,10 & 1\,\,1 \\
\end{tabular}
\end{ruledtabular}

\label{tab:H2_STO3G_restricted_mapping}
\end{table}

\begin{table*}[t]
\centering
\caption{Mapping of excitation operators to Pauli operators for a total-spin-restricted $\ch{H2}$ molecule in the STO-3G basis set. Note that $I_k$ in the Entry-operator column denotes an identity operator on the $k$-th qubit.}

\begin{ruledtabular}
\begin{tabular}{ccccc}
Excitation & fermionic basis & Qubit state basis & \textrm{Entry-operator} & Pauli \\
\hline
$E_{10}$ & $\ket{0110}_{\text{f}}\bra{0101}_{\text{f}}+\ket{1010}_{\text{f}}\bra{1001}_{\text{f}}$ & $\ket{01}_{\text{q}}\bra{00}_{\text{q}}+\ket{11}_{\text{q}}\bra{10}_{\text{q}}$ & $I_1Q^+_0$ & $\frac{1}{2}IX-\frac{i}{2}IY$ \\

$E_{32}$ & $\ket{1001}_{\text{f}}\bra{0101}_{\text{f}}+\ket{1010}_{\text{f}}\bra{0110}_{\text{f}}$ & $\ket{10}_{\text{q}}\bra{00}_{\text{q}}+\ket{11}_{\text{q}}\bra{01}_{\text{q}}$ & $Q^+_1I_0$ & $\frac{1}{2}XI-\frac{i}{2}YI$ \\

$E_{00}$ & $\ket{0101}_{\text{f}}\bra{0101}_{\text{f}}+\ket{1001}_{\text{f}}\bra{1001}_{\text{f}}$ & $\ket{00}_{\text{q}}\bra{00}_{\text{q}}+\ket{10}_{\text{q}}\bra{10}_{\text{q}}$ & $I_1N^{(0)}_0$ & $\frac{1}{2}II+\frac{1}{2}IZ$ \\

$E_{11}$ & $\ket{0110}_{\text{f}}\bra{0110}_{\text{f}}+\ket{1010}_{\text{f}}\bra{1010}_{\text{f}}$ & $\ket{01}_{\text{q}}\bra{01}_{\text{q}}+\ket{11}_{\text{q}}\bra{11}_{\text{q}}$ & $I_1N^{(1)}_0$ & $\frac{1}{2}II-\frac{1}{2}IZ$ \\

$E_{22}$ & $\ket{0101}_{\text{f}}\bra{0101}_{\text{f}}+\ket{0110}_{\text{f}}\bra{0110}_{\text{f}}$ & $\ket{00}_{\text{q}}\bra{00}_{\text{q}}+\ket{01}_{\text{q}}\bra{01}_{\text{q}}$ & $N^{(0)}_1I_0$ & $\frac{1}{2}II+\frac{1}{2}ZI$ \\

$E_{33}$ & $\ket{1001}_{\text{f}}\bra{1001}_{\text{f}}+\ket{1010}_{\text{f}}\bra{1010}_{\text{f}}$ & $\ket{10}_{\text{q}}\bra{10}_{\text{q}}+\ket{11}_{\text{q}}\bra{11}_{\text{q}}$ & $N^{(1)}_1I_0$ & $\frac{1}{2}II-\frac{1}{2}ZI$ \\
\end{tabular}
\end{ruledtabular}

\label{tab:stepbystep_mapping}
\end{table*}

With the mapping from the fermionic configurations to the qubit states and Eq. (\ref{eq: Epq as T}), we can transform the excitation operators in the fermionic basis to the qubit state basis where the entry-operators can help us build the Pauli operators as shown in Table \ref{tab:stepbystep_mapping}.

Lastly, with the help of the identity in Eq. (\ref{eq:double_excitation_identity}) that represents double excitation terms with the excitation operators $E_{pq}$, we can write the second-quantized Hamiltonian in Eq. (\ref{eq:second_hamil}) as a qubit Hamiltonian. For example, at the interatomic distance of 0.735 Å (the equilibrium distance in the STO-3G basis set) for the two hydrogen atoms, the qubit Hamiltonian can be written as 

\begin{align}
\begin{split}
  H_{\text{q}} = 
    &-1.052373 \cdot I_1I_0 -0.397937 \cdot Z_1I_0 \\ &-0.397937 \cdot I_1Z_0 +0.011280 \cdot Z_1Z_0 \\ &+0.180931 \cdot X_1X_0.
\end{split}
\end{align}

\subsubsection{$\ch{H2}$, STO-3G, total-spin-unrestricted}
In this subsection, we do not restrict the total spin to singlet configurations, so there will be triplet electronic configurations present in the system. 
Similarly, for an $\ch{H2}$ molecule in the STO-3G basis set, we can write an electronic configuration in the fermionic occupation basis as
\begin{equation}
    \ket{\psi} 
  = \ket{f_{\sigma_{1s, u\downarrow}}, f_{\sigma_{1s, u\uparrow}}, f_{\sigma_{1s, g\downarrow}}, f_{\sigma_{1s, g\uparrow}}}
  = \ket{f_3, f_2, f_1, f_0}.
  \label{f_unrestricted}
\end{equation}
Note that the ordering of occupation numbers in Eq.~(\ref{f_unrestricted}) is slightly different from the total-spin-restricted case of  Eq.~(\ref{f_restricted}) in Sec.~\ref{subsubsec: H2_total_spin_restricted}.
However, there are 6 fermionic configurations in this case so 3 qubits are required to represent all these configurations as shown in Table \ref{tab:H2_STO3G_unrestricted_mapping} (also in an ascending order).

\begin{table}[hbt!]
\centering
\caption{Mapping of fermionic configurations to qubit basis state for a total-spin-unrestricted $\ch{H2}$ molecule in the STO-3G basis set.}
\begin{ruledtabular}
\begin{tabular}{ccc}
Filled spin-orbitals&$f_3f_2\,f_1f_0$&$q_2\,q_1\,q_0$\\
\hline
$\sigma_{1s, g\downarrow}\sigma_{1s, g\uparrow}$   & 0011 & 000 \\
$\sigma_{1s, u\uparrow}\sigma_{1s, g\uparrow}$     & 0101 & 001 \\
$\sigma_{1s, u\uparrow}\sigma_{1s, g\downarrow}$   & 0110 & 010 \\
$\sigma_{1s, u\downarrow}\sigma_{1s, g\uparrow}$   & 1001 & 011 \\
$\sigma_{1s, u\downarrow}\sigma_{1s, g\downarrow}$ & 1010 & 100 \\
$\sigma_{1s, u\downarrow}\sigma_{1s, u\uparrow}$   & 1100 & 101 \\
\end{tabular}
\end{ruledtabular}

\label{tab:H2_STO3G_unrestricted_mapping}
\end{table}
Thus, using similar mapping procedure shown in the previous example, we can write the qubit Hamiltonian at the interatomic distance of hydrogen bond length 0.735 Å as
\begin{align}
\begin{split}
  H_{\text{q}} = 
    &-0.837333 \cdot I_2I_1I_0 -0.198969 \cdot I_2I_1Z_0 \\
    &-0.305506 \cdot I_2Z_1I_0 -0.198969 \cdot I_2Z_1Z_0 \\
    &-0.464882 \cdot Z_2I_1I_0 +0.050873 \cdot Z_2I_1Z_0 \\
    &+0.066945 \cdot Z_2Z_1I_0 +0.050873 \cdot Z_2Z_1Z_0 \\
    &-0.045233 \cdot I_2I_1X_0 +0.045233 \cdot I_2Z_1X_0 \\
    &-0.045233 \cdot Z_2I_1X_0 +0.045233 \cdot Z_2Z_1X_0 \\
    &-0.045233 \cdot X_2I_1X_0 -0.045233 \cdot X_2Z_1X_0 \\
    &+0.045233 \cdot Y_2I_1Y_0 +0.045233 \cdot Y_2Z_1Y_0.
\end{split}
\end{align}

\subsubsection{$\ch{H2}$, 6-31G, Total-spin-restricted} \label{sec: H2_631G_illustrative}
By adopting symmetries in conservation of electrons and spins ($\mathbb{Z}_2$ symmetries, two qubits reduction), it is possible for the parity or BK mappings to reduce 2 qubits for the encoding of a system with an arbitrary number of spin-orbitals \cite{bravyi_tapering_2017}.
For the case of an $\ch{H2}$ molecule in the STO-3G basis set, our QEE method as well as the parity and BK mappings only require 2 qubits, i.e., a reduction of 2 qubits from a 4-qubit setting.
Therefore, it is more pedagogical to do experiments on a larger system where the QEE reduces more than 2 qubits. 
We show here the reduction of 4 qubits in the case of $\ch{H2}$ in the 6-31G basis set, where there are 8 spin-orbitals and 2 electrons in the system.
The VQE data is  shown in Sec.~\ref{sec: result}.

Similar to what we have done for the case of the STO-3G basis set, for an $\ch{H2}$ molecule in the 6-31G basis set, we can write an electronic configuration in the fermionic occupation basis as
\begin{widetext}
\begin{align}
 \left |
  f_{\sigma_{2s, u\downarrow}}, f_{\sigma_{2s, g\downarrow}}, f_{\sigma_{1s, u\downarrow}}, f_{\sigma_{1s, g\downarrow}},
  f_{\sigma_{2s, u\uparrow}},
  f_{\sigma_{2s, g\uparrow}},
  f_{\sigma_{1s, u\uparrow}},
  f_{\sigma_{1s, g\uparrow}}
     \right \rangle
     = \ket{f_7, f_6, f_5, f_4, f_3, f_2, f_1, f_0},
\end{align}
\end{widetext}
where the 2s atomic orbitals are also included to form molecular orbitals. 
With the electron number being conserved and the total spin being restricted such that there are only singlet electronic configurations, there are 16 fermionic configurations, so only 4 ($\log_216$) qubits are used to map these configurations to the qubit basis states as shown in Table \ref{tab:H2_631G_restricted_mapping}. 
Thus, we can simulate this system with 4 qubits using QEE, and a qubit Hamiltonian can be constructed in the same way as previous examples (see Appendix \ref{appendix:H2_0.745_QEE_Hamiltonian} for the qubit Hamiltonian). 
We show the potential energy surface of this system at different interatomic distances of the two hydrogen atoms in Sec.~\ref{sec: result} by running the VQE algorithms for the QEE qubit Hamiltonian.

\begin{table}[hbt!]
\centering
\caption{Mapping of fermionic configurations to qubit basis state for a total-spin-restricted $\ch{H2}$ molecule in the 6-31G basis set.}
\begin{ruledtabular}
\begin{tabular}{ccc}
Filled spin-orbitals & $f_7f_6f_5f_4\,f_3f_2f_1f_0$ & $q_3q_2\,q_1q_0$ \\
\hline
$\sigma_{1s, g\downarrow}\sigma_{1s, g\uparrow}$ & 0001\,\,0001 & 00\,\,00 \\
$\sigma_{1s, g\downarrow}\sigma_{1s, u\uparrow}$ & 0001\,\,0010 & 00\,\,01 \\
$\sigma_{1s, g\downarrow}\sigma_{2s, g\uparrow}$ & 0001\,\,0100 & 00\,\,10 \\
$\sigma_{1s, g\downarrow}\sigma_{2s, u\uparrow}$ & 0001\,\,1000 & 00\,\,11 \\
$\sigma_{1s, u\downarrow}\sigma_{1s, g\uparrow}$ & 0010\,\,0001 & 01\,\,00 \\
$\sigma_{1s, u\downarrow}\sigma_{1s, u\uparrow}$ & 0010\,\,0010 & 01\,\,01 \\
$\sigma_{1s, u\downarrow}\sigma_{2s, g\uparrow}$ & 0010\,\,0100 & 01\,\,10 \\
$\sigma_{1s, u\downarrow}\sigma_{2s, u\uparrow}$ & 0010\,\,1000 & 01\,\,11 \\
$\sigma_{2s, g\downarrow}\sigma_{1s, g\uparrow}$ & 0100\,\,0001 & 10\,\,00 \\
$\sigma_{2s, g\downarrow}\sigma_{1s, u\uparrow}$ & 0100\,\,0010 & 10\,\,01 \\
$\sigma_{2s, g\downarrow}\sigma_{2s, g\uparrow}$ & 0100\,\,0100 & 10\,\,10 \\
$\sigma_{2s, g\downarrow}\sigma_{2s, u\uparrow}$ & 0100\,\,1000 & 10\,\,11 \\
$\sigma_{2s, u\downarrow}\sigma_{1s, g\uparrow}$ & 1000\,\,0001 & 11\,\,00 \\
$\sigma_{2s, u\downarrow}\sigma_{1s, u\uparrow}$ & 1000\,\,0010 & 11\,\,01 \\
$\sigma_{2s, u\downarrow}\sigma_{2s, g\uparrow}$ & 1000\,\,0100 & 11\,\,10 \\
$\sigma_{2s, u\downarrow}\sigma_{2s, u\uparrow}$ & 1000\,\,1000 & 11\,\,11 \\
\end{tabular}
\end{ruledtabular}
\label{tab:H2_631G_restricted_mapping}
\end{table}

\subsubsection{$\ch{LiH}$, STO-3G, Total-spin-restricted} \label{sec: LiH_STO3G_illustrative}
We also apply our encoding method to a $\ch{LiH}$ molecule in the STO-3G basis set with the core frozen and the $\text{2p}_{y}$ orbital removed (assuming zero filling) for the Li atom. In this case, there are also 8 spin-orbitals and 2 electrons in the system so it requires 8 qubits for the JW encoding. We choose to only consider the singlet configurations so this system is similar to the case in Sec.~\ref{sec: H2_631G_illustrative}. Thus, this system requires 4 qubits for our encoding scheme since there are 16 fermionic configurations. We show the potential energy surfaces of this system in Sec.~\ref{sec: result} as well.  

Note that we present the reduction using particle-conserving and total-spin symmetries, but
our method has the flexibility of removing any configurations
by perturbation arguments or imposing other symmetries or constraints
to systematically reduce the dimension of a Hilbert space.

\section{RESULTS AND DISCUSSION} \label{sec: result}

\begin{figure*}[t]

\begin{subfigure}{0.45\textwidth}
\includegraphics[width=\textwidth]{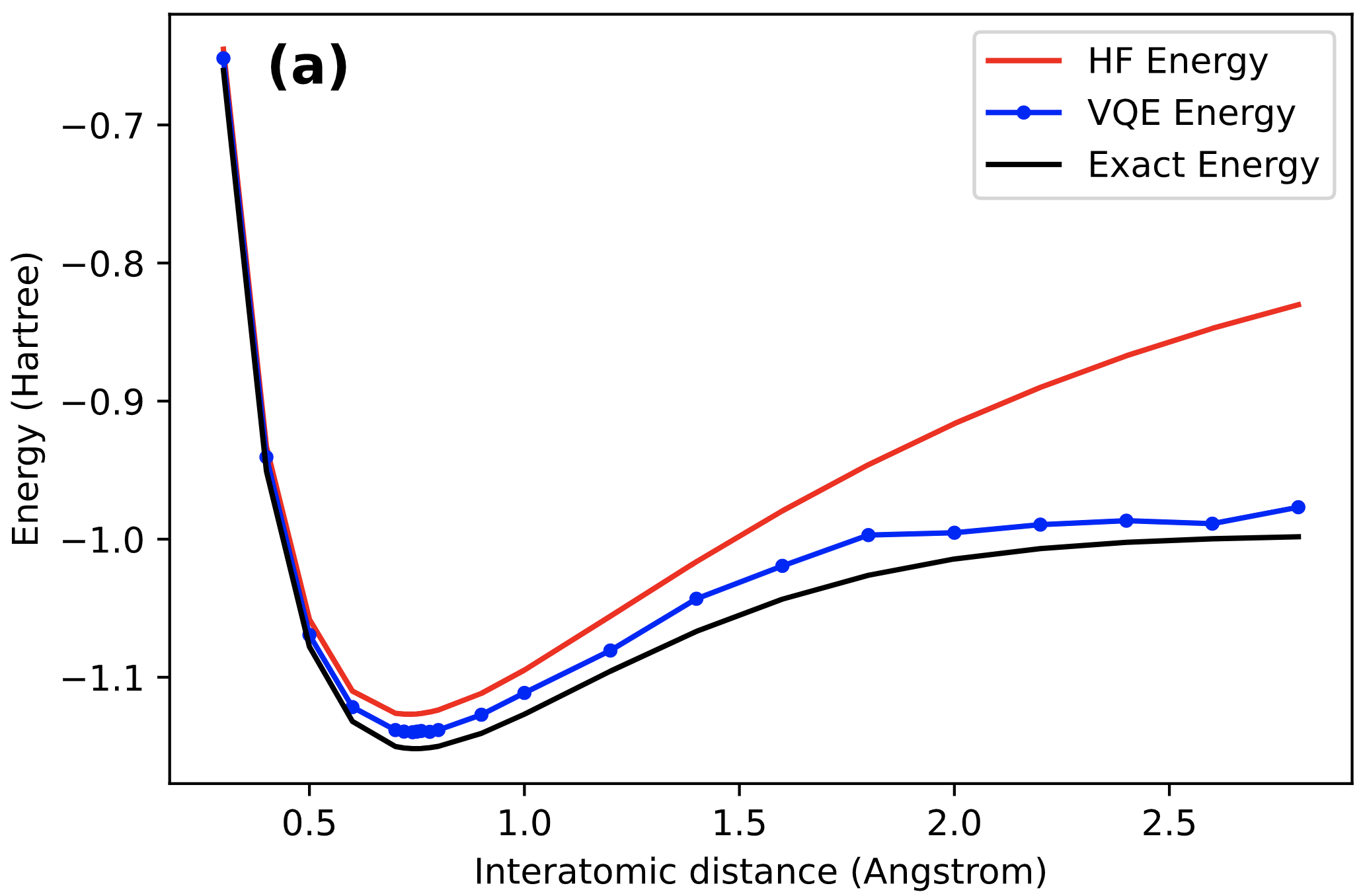} 
\phantomsubcaption
\label{fig:H2_VQE_10_PES}
\end{subfigure}
\begin{subfigure}{0.45\textwidth}
\includegraphics[width=\textwidth]{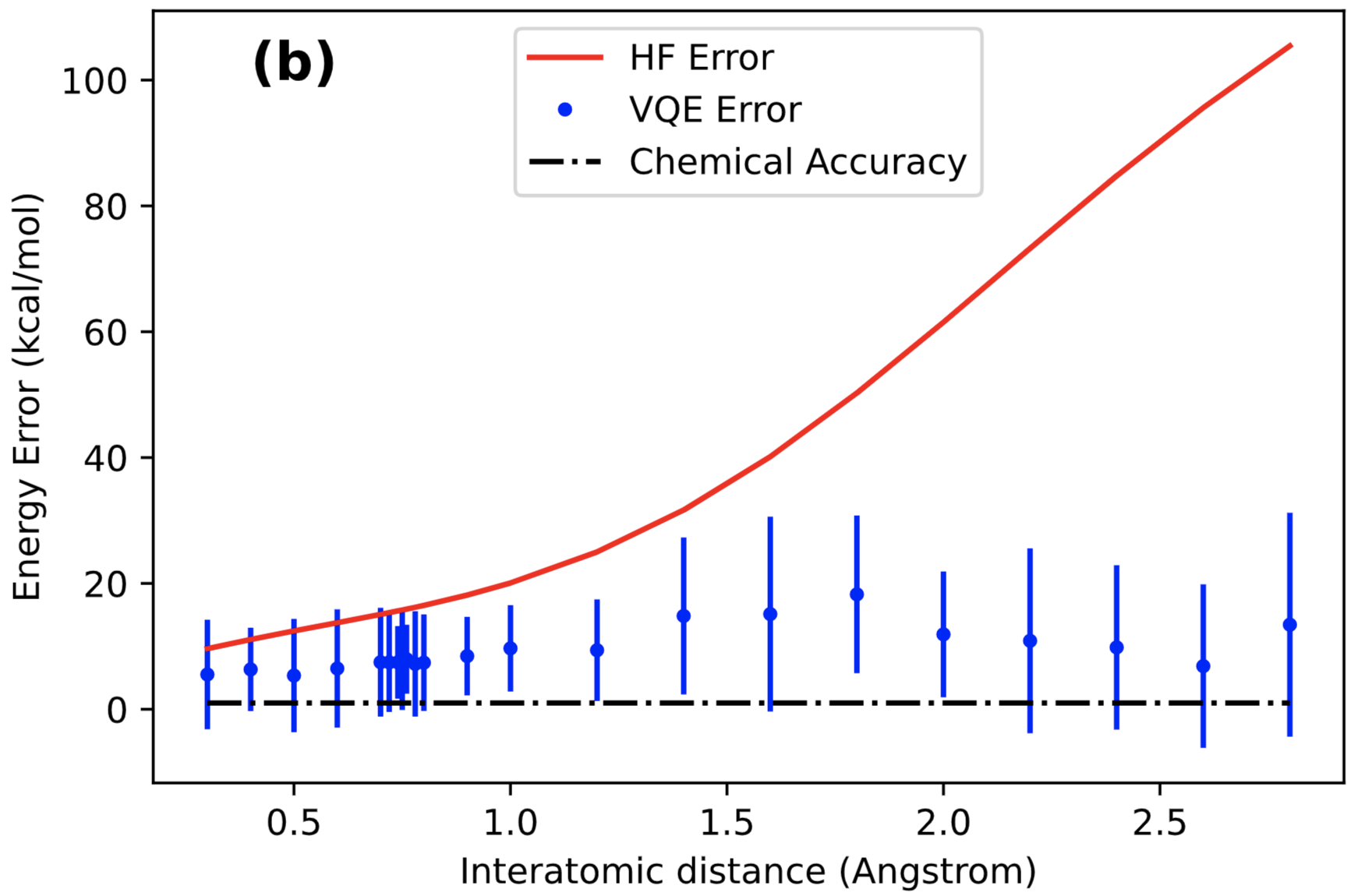}
\phantomsubcaption
\label{fig:H2_VQE_10_Error}
\end{subfigure}

\begin{subfigure}{0.45\textwidth}
\includegraphics[width=\textwidth]{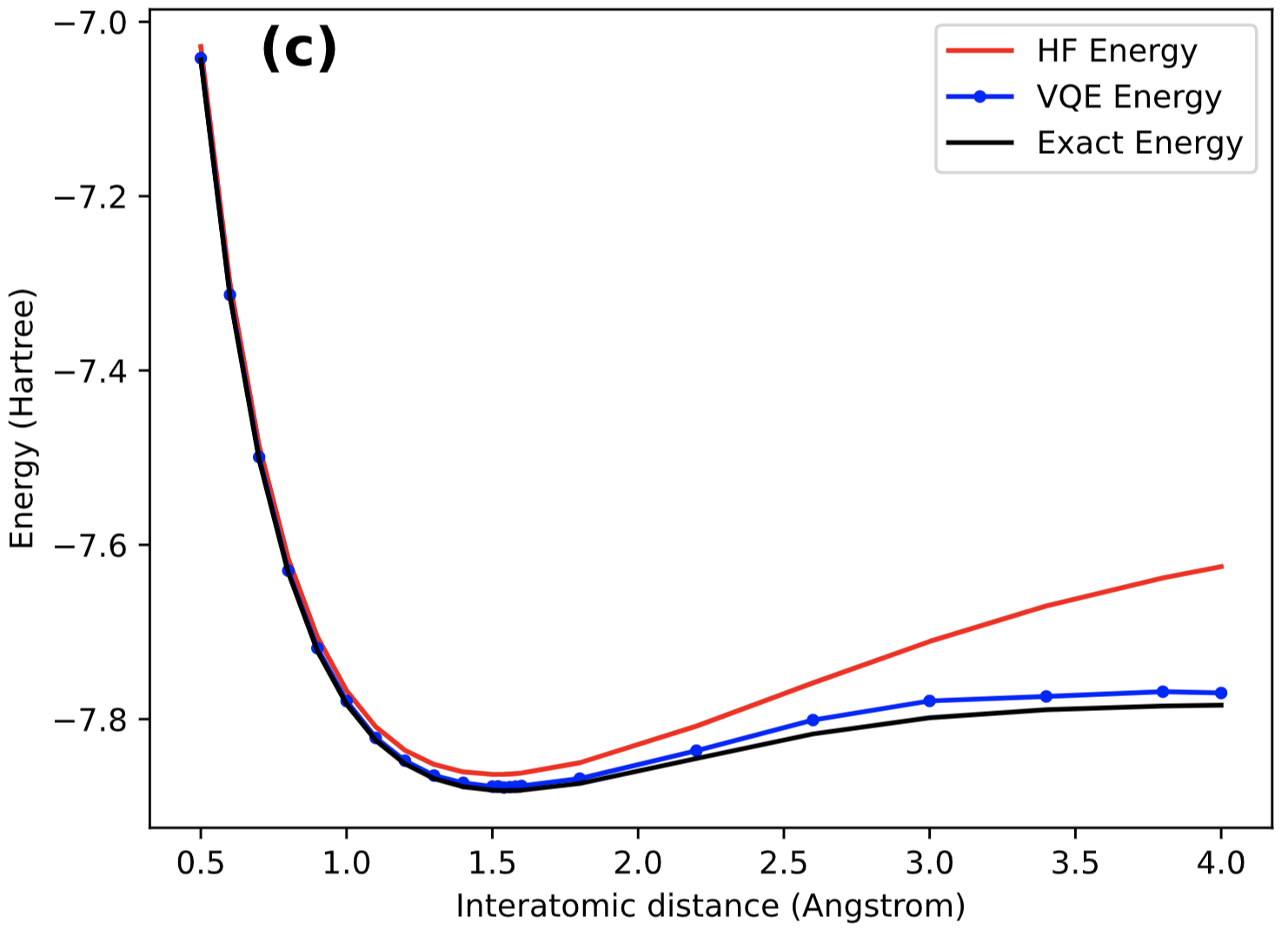} 
\phantomsubcaption
\label{fig:LiH_VQE_10_PES}
\end{subfigure}
\begin{subfigure}{0.45\textwidth}
\includegraphics[width=\textwidth]{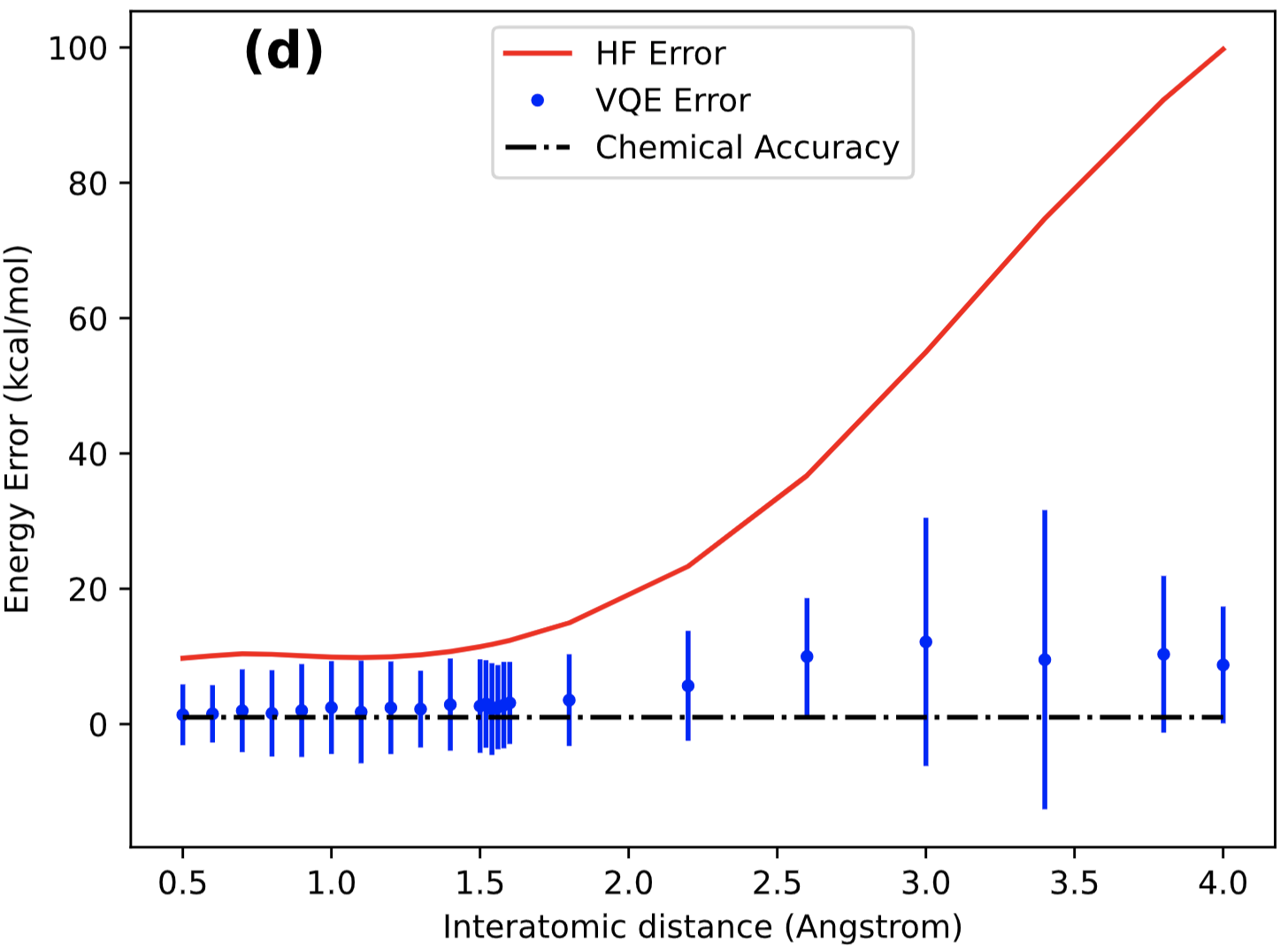}
\phantomsubcaption
\label{fig:LiH_VQE_10_Error}
\end{subfigure}

\caption{Potential energy surfaces and energy errors of an $\ch{H2}$ molecule in the 6-31G basis set and a $\ch{LiH}$ molecule in the STO-3G basis set. (a) $\ch{H2}$ Potential Energy Surfaces. (b) $\ch{H2}$ Energy Errors. (c) $\ch{LiH}$ Potential Energy Surfaces. (d) $\ch{LiH}$ Energy Errors. For (a) and (c),  The red curves were obtained using the Hartree--Fock method and the black curves represent the exact energy surfaces obtained by diagonalization of the JW qubit Hamiltonian with classical algorithms. The blue curves were the extrapolated energies over three different CNOT gate counts with the energies of each CNOT gate count being averaged over 10 sets of VQE experiments and with $10^4$ shots for each iteration using QEE qubit Hamiltonian. The simulations were performed on the QASM simulator with a noise model implemented from the calibration data of IBM Quantum device \texttt{ibmq\_santiago} shown in Appendix \ref{appendix:calibration} using qubit the 0, 1, 2, and 3. For (b) and (d), the curves and data points represent the energy errors with respect to the exact energy surfaces of each molecule. The error bars represent the 95\% confidence uncertainties that took the standard deviations of experiments at each CNOT gate count and the residues of linear regression into account. Most of the distributions agree within the chemical accuracy defined as 1 kcal/mol.}
\label{fig:PES_Error}
\end{figure*}

Simulations for an $\ch{H2}$ molecule in the 6-31G basis set and a $\ch{LiH}$ molecule in the 
STO-3G basis set with restricted total spin at some given interatomic distances of the two atoms of each molecule are performed with the quantum subroutine of the VQE algorithms
using Qiskit. The classical optimization of the variational parameters was done using the Constrained Optimization by Linear Approximation (COBYLA) algorithm with a maximum iteration number of 500. The QEE Hamiltonian of the system at different interatomic distances are generated using the implementation method (code) at \cite{tsai_qubit-efficient_2021}. The exact energies and ground-state configurations of the JW encoded qubit Hamiltonian are used to verify the correctness of these QEE qubit Hamiltonian using the statevector simulator in Qiskit. 

The ansatz circuits used in the VQE algorithms are 
the real-amplitudes 2-local circuits \cite{noauthor_qiskitcircuitlibraryrealamplitudes_nodate} with two layers of linear entanglement (Fig.~\ref{fig:AnsatzCircuit6}) and its variants with redundant CNOT pairs (Fig.~\ref{fig:AnsatzCircuit12} and Fig.~\ref{fig:AnsatzCircuit18}) so the CNOT gate counts of the circuits were 6, 12, and 18, respectively. To mimic the effect of realistic noisy quantum machine, the simulations are performed on the QASM simulator with a noise model implemented from the calibration data of IBM Quantum device \texttt{ibmq\_santiago} shown in Appendix \ref{appendix:calibration}. The first 4 qubits of the device \texttt{ibmq\_santiago} in the noise model are used because they form a linear chain of qubits so that no extra CNOT gates were needed to implement our ansatz circuits. The experiments are run with $10^4$ or $10^5$ readout shots per circuit with the measurement error being mitigated by the inverse of a calibrated matrix. For each CNOT gate count, 10 independent VQE experiments are done to find the error-free limit of the energy expectation values using the linear extrapolating error mitigating method.

The potential energy surfaces of the $\ch{H2}$ system are plotted in Fig.~\ref{fig:H2_VQE_10_PES} from discrete data points at the interatomic distances of 0.3 to 2.8 angstrom with more data points plotted from 0.7 to 0.8 angstrom close to the equilibrium distance, and the potential energy surfaces of the $\ch{LiH}$ system are plotted in Fig.~\ref{fig:LiH_VQE_10_PES} from 0.5 to 4.0 angstrom with more data points plotted from 1.5 to 1.6.
The number of measurement shots for both cases is $10^4$
The Hartree--Fock energies are obtained using the ``PySCF'' Python-based chemical simulation package and the exact energies (ground-state energies) are obtained from diagonalization of the JW encoded qubit Hamiltonian. Since there are errors arisen from finite numbers of measurements (shots), SPAM errors of the qubits, and single-qubit gate errors in VQE experiments even at the error-free limits of the CNOT gates, 10 independent VQE experiments are done for each of the CNOT gate counts. The errors of the extrapolated energies with respect to the exact energies are plotted in Fig.~\ref{fig:H2_VQE_10_Error} and in Fig.~\ref{fig:LiH_VQE_10_Error} with the error bars showing the uncertainties in 95\% confidence intervals. These uncertainties are propagated from the residues of the linear regressions and the standard deviations of the distributions at each of the CNOT gate counts. Even though the energy errors of the VQE experiments are slightly higher than the chemical accuracy (1 kcal/mol), most of the uncertainties
reach the chemical accuracy. Thus, another experiment on increasing the measurement shots per circuit is done to further test the feasibility of this method in the NISQ era. We will discuss this result later.

\begin{figure*}[t]
    \begin{subfigure}{0.45\textwidth}
    \includegraphics[width=\textwidth]{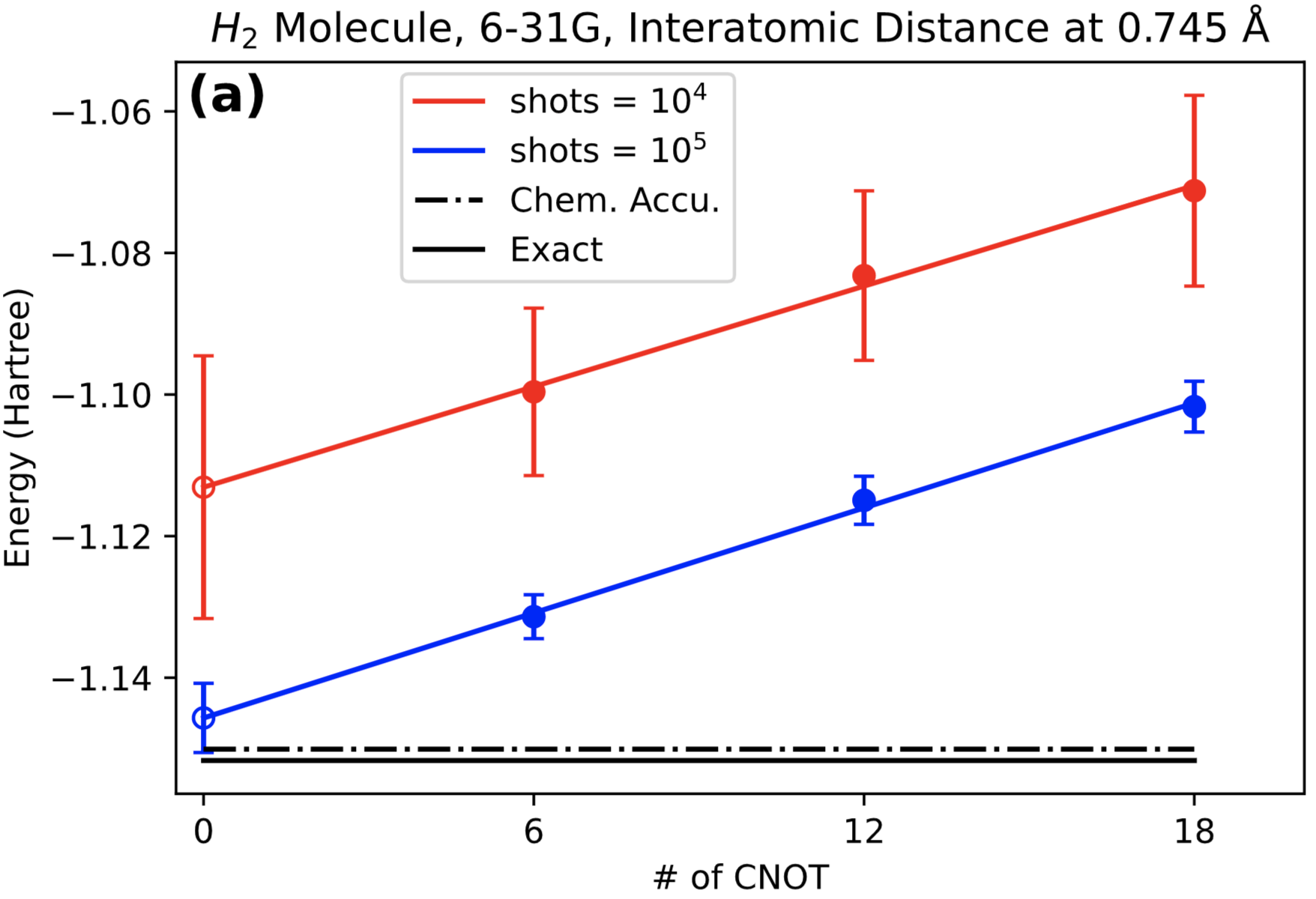} 
    \phantomsubcaption
    \label{fig:H2_0.745_No_VQE}
    \end{subfigure}
    \begin{subfigure}{0.45\textwidth}
    \includegraphics[width=\textwidth]{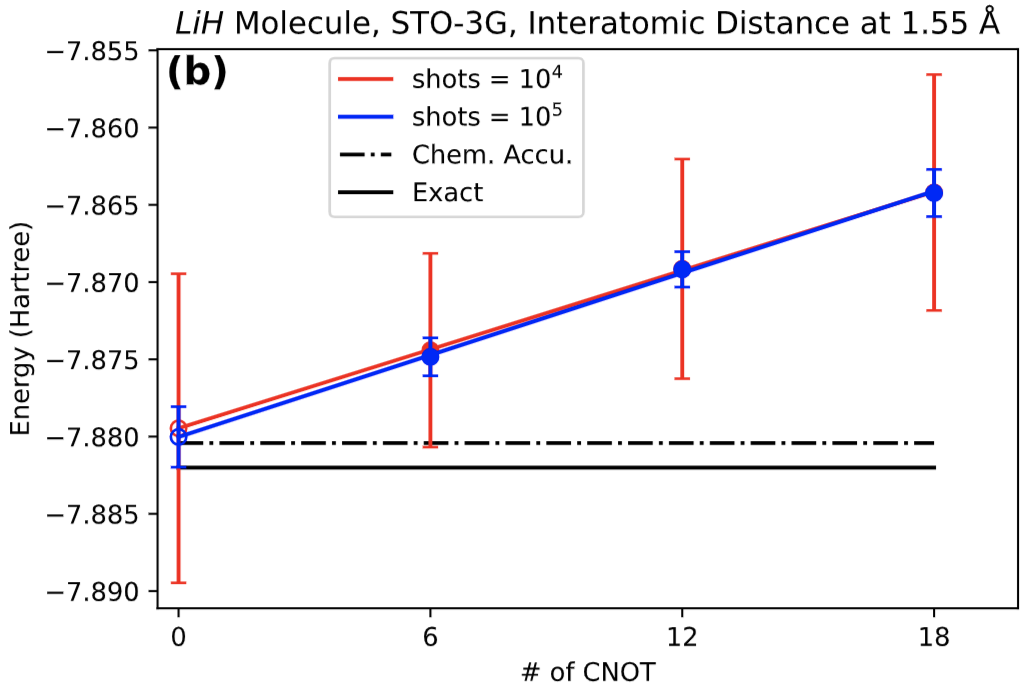}
    \phantomsubcaption
    \label{fig:LiH_1.55_No_VQE}
    \end{subfigure}
    \caption{The extrapolated energies to the error-free limits of an $\ch{H2}$ molecule in the 6-31G basis set at an interatomic distance of 0.745 Å and a $\ch{LiH}$ molecule in the STO-3G basis set at an interatomic distance of 1.55 Å
      with the experiments run on a noisy simulator. (a) $\ch{H2}$ Extrapolated Energies. (b) $\ch{LiH}$ Extrapolated Energies.
The simulations were performed on the QASM simulator with a noise model implemented from the calibration data of IBM Quantum device \texttt{ibmq\_santiago} shown in Appendix \ref{appendix:calibration} using qubits 0, 1, 2, and 3.
      Each of the data points in solid circles was averaged over 10 sets of experiments using $10^4$ or $10^5$ shots per circuit with the rotation angles pre-evaluated on a noiseless simulator. Each of the error bars of the solid circles is twice of the standard deviation over the energy distribution for the 10 experiments. The red and blue lines are the linear fits of the energies in solid circles. The hollow circles are the extrapolated energies with the error bars being the 95\% confidence uncertainties estimated from the standard deviations and linear regression residues.
The distributions of the extrapolated energies using $10^5$ shots agree within the chemical accuracy and there is significantly less variance in this case. Note that the blue circles in (b) are slightly lower than the red circles.}
    \label{fig:No_VQE}
\end{figure*}

In Kandala et al.'s work \cite{kandala_error_2019}, they used a similar extrapolation method to estimate the ground-state energies of an $\ch{H2}$ molecule and a $\ch{LiH}$ molecule in the STO-3G basis set using the BK encoding. Instead of stacking up redundant CNOT gates, they extended the pulse times of the CNOT gates to amplify the major errors. In general, our ground-state potential energy surfaces obtained from VQE have less error than their results at different interatomic distances. Since, for the $\ch{H2}$ cases, we used a larger basis set (6-31G) and more qubits than those of their work, it would be more reasonable to compare the $\ch{LiH}$ cases. The authors in \cite{kandala_error_2019} also froze the core orbital of $\ch{LiH}$ but they remove both the $\text{2p}_{y}$ and $\text{2p}_{z}$ orbitals. Thus, there are 2 extra spin-orbitals in our system, but we use the same number of qubits (4 qubits) as theirs. With these 2 extra spin-orbitals, we can capture slightly more correlation using the same number of qubits. It can be observed from Fig.~\ref{fig:LiH_VQE_10_Error} and the bottom right of Fig. 4. in \cite{kandala_error_2019} that some of their extrapolated energies were found to be quite far away from chemical accuracy comparing to our result. Such improvement was not directly due to the inclusion of the $\text{2p}_{z}$ orbital in our system, because the $\text{2p}_{z}$ orbital does not interact strongly with other orbitals, and the energy errors are the comparison criteria for the VQE performance here. Instead, the entire Hilbert space is spanned by the configurations with desired symmetries in our case, so the final state would not include undesired or unphysical configurations after an optimization search using a heuristic hardware-efficient ansatz circuit. This shows that we can use the same number of qubits to reach a better accuracy for the same chemical system. 

\begin{figure*}[t]
    \begin{subfigure}{0.45\textwidth}
    \includegraphics[width=\textwidth]{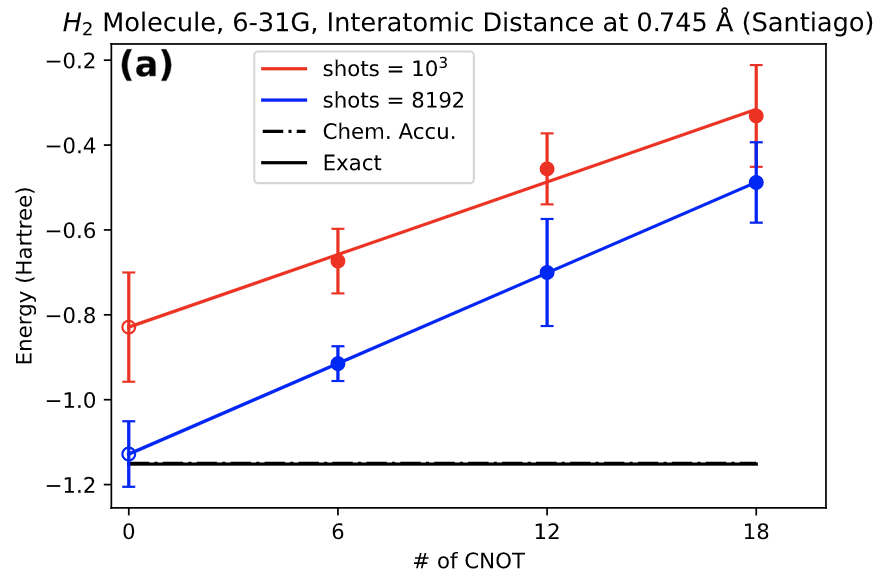} 
    \phantomsubcaption
    \label{fig:H2_0.745_No_VQE_real_device}
    \end{subfigure}
    \begin{subfigure}{0.45\textwidth}
    \includegraphics[width=\textwidth]{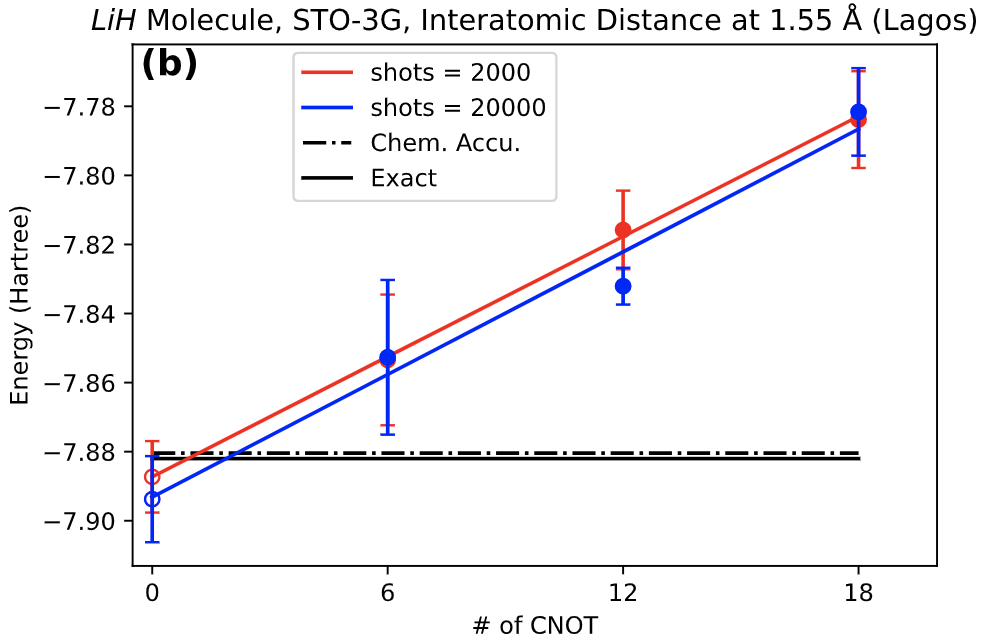}
    \phantomsubcaption
    \label{fig:LiH_1.55_No_VQE_real_device}
    \end{subfigure}
    \caption{ The extrapolated energies to the error-free limits of an $\ch{H2}$ molecule in the 6-31G basis set at an interatomic distance of 0.745 Å and a $\ch{LiH}$ molecule in the STO-3G basis set at an interatomic distance of 1.55 Å with the experiments run on quantum devices. (a) $\ch{H2}$ Extrapolated Energies. The experiments were performed on qubits 0, 1, 2, and 3 of \texttt{ibmq\_santiago} on October 13, 2021. (b) $\ch{LiH}$ Extrapolated Energies. The experiments were performed on qubits 0, 1, 3, and 5 of \texttt{ibmq\_lagos} on December 14, 2021. Each of the data points in solid circles was averaged over 10 sets of experiments using 1000 or 8192, 2000 or 20000 shots per circuit with the rotation angles pre-evaluated on a noiseless simulator. Each of the error bars of the solid circles is twice of the standard deviation over the energy distribution for the 10 experiments. The red and blue lines are the linear fits of the energies in solid circles. The hollow circles are the extrapolated energies with the error bars being the 95\% confidence uncertainties estimated from the standard deviations and linear regression residues.}
\label{fig:No_VQE_real_device}
\end{figure*}

To demonstrate that chemical accuracy could be achievable for QEE on the present and near-term devices, more measurement shots per circuit are performed on the same noise model to investigate how noise in a finite number of measurements affects the variational computations and the results are plotted in Fig.~\ref{fig:No_VQE}. The tunable parameters (the rotation angles) of the ansatz circuits are first optimized and obtained using Qiskit's statevector simulator (noiseless simulator) so the parameters (rotation angles) of the ansatz circuits are fixed in this experiment to prevent the uncertainties in the variational process. The measurement error mitigation and extrapolation to the error-free limit are also done in this experiment. Note that the error distribution of the experiment using $10^4$ shots per iteration in Fig.~\ref{fig:H2_0.745_No_VQE} is far from chemical accuracy because the parameters are pre-optimized on a noiseless simulator and fixed during the experiments, so the robustness from VQE against some noise, such as noise in single-qubit gates, would not be present as compared to the experiments in Fig.~\ref{fig:PES_Error} \cite{mcclean_theory_2016, omalley_scalable_2016}. Figure \ref{fig:H2_0.745_No_VQE} shows that the error of the extrapolated energy with respect to the exact ground-state energy obtained by Hamiltonian diagonalization at the interatomic distance of 0.745 Å (the equilibrium distance in the 6-31G basis set) decreases as the number of shots per circuit increases and the uncertainties decreases significantly as well.
The distribution of the extrapolated energy using $10^5$ shots at the error-free limit agrees with the exact ground-state energy within chemical accuracy.
A similar result is shown for a $\ch{LiH}$ molecule in Fig.~\ref{fig:LiH_1.55_No_VQE} where both the error and the uncertainty of the extrapolated energy decrease using more shots per circuit.
In particular, the overestimated extrapolated values of the ground-state energy lower than the exact energy obtained by diagonization indicated by the error bar (red solid line) at the error-free point in  the case of $10^4$ shots is significantly improved (see the error bar in blue solid line at the error-free point) by using $10^5$ shots.  
Similar trends could be seen in  Fig.~\ref{fig:No_VQE_real_device} where the experiments were done on IBM quantum devices.
These results indicate that substantial numbers of measurements are still needed for the simulations on noisy quantum devices.
Therefore, given that the 
measurement error mitigation and extrapolation to the error-free limit for two-qubit CNOT gates are performed, the main concern for our method for the $\ch{H2}$ molecule in the 6-31G basis set and  the $\ch{LiH}$ molecule in STO-3G basis set is the probabilistic behavior of a finite number of measurements in the noisy simulations
instead of decoherence or SPAM error of the quantum processors.
In short, all of our results demonstrate that quantum chemical simulations using QEE qubit Hamiltonians are suitable in the NISQ era. Some additional admissible use cases for our QEE scheme are provided in Table \ref{tab:JW_QEE_Terms_Comparison}.

\begin{table*}[t]
\centering
\caption{Comparison between qubit counts and Hamiltonian Pauli term counts of the JW encoding and QEE for certain molecules (in the STO-3G basis sets and equilibrium bond distances) with some molecular orbitals being frozen/removed. Note that the molecular orbitals are ordered from the lowest to the highest energies and the labels start from 0.}

\begin{ruledtabular}
\begin{tabular}{cccccc}
Molecule & Frozen/Removed Orbitals & JW Qubit Count & JW Terms & QEE Qubit Count & QEE Terms \\
\hline
$\ch{LiH} $ & 0, 3  & 8 & 193      & 4         & 100      \\
$\ch{HF}  $ & N/A  & 12 & 631      & 6         & 1184      \\
$\ch{HF}  $ & 0    & 10 & 276      & 6         & 608       \\
$\ch{HCl} $ & 0    & 18 & 3772     & 8         & 8960      \\
$\ch{HCl} $ & 0, 1  & 16 & 2329     & 6         & 640       \\
$\ch{HBr} $ & 0-2  & 32 & 40705    & 8         & 18490     \\
$\ch{HBr} $ & 0-4  & 28 & 21891    & 8         & 18472     \\
$\ch{F2}  $  & 0, 1  & 16 & 1177     & 6         & 1040      \\
$\ch{Cl2} $ & 0, 1  & 32 & 21481    & 8         & 17500     \\
$\ch{Cl2} $ & 0-9  & 16 & 1177     & 6         & 1040      \\
$\ch{Br2} $ & 0-27 & 16 & 1177     & 6         & 1040      \\
$\ch{I2}  $  & 0-45 & 16 & 1177     & 6         & 1040     
\end{tabular}
\end{ruledtabular}

\label{tab:JW_QEE_Terms_Comparison}
\end{table*}

\section{CONCLUSION} \label{sec: conclusion}
In conclusion, in order to realize molecular quantum simulations of large systems, quantum hardware should be improved and the quantum algorithms should also be optimized to respect hardware constraints. 
We have designed an alternative encoding scheme to reduce qubit resources by filling up (most of) the Hilbert space with desired configurations. 
For an $\ch{H2}$ molecule in the 6-31G basis set and a $\ch{LiH}$ molecule in the STO-3G basis set, we have reduced the required number of qubits from 8 to 4 with our encoding scheme and have simulated the molecules by incorporating a noise model from a real IBM Quantum machine 
with the distribution of the extrapolated energies agreeing with the exact energies obtained by Hamiltonian diagonalization to chemical accuracy.

For \revise{electron number $m \leq\frac{N}{2}$}, the number of qubits $Q$ needed  in our QEE scheme \revise{has an upper bound of $\mathcal O(m\log_2N)$, and for $m > \frac{N}{2}$, the upper bound of the number of qubits $Q$ is of $\mathcal O((N-m)\log_2N)$} (see Appendix \ref{appendix:Q_scaling} for the derivation), which is a noticeable advantage compared to $Q=N$ in most present encoding schemes. 
For a given molecular system, the number of spin-orbitals $N$ for quantum chemical simulation increases rapidly as the basis set becomes larger. 
For example, a water molecule with the ccpV5Z basis set has 201 molecular orbitals, or 402 spin-orbitals. Using the parity or BK mapping with $\mathbb Z_2$ reduction, one still needs 400 qubits, which may be beyond the capability of quantum processors in the near future. 
With QEE, we use only $Q=\left\lceil\log_2{402\choose 10}\right\rceil = 65$ qubits in the total-spin-unrestricted case, which is a prominent reduction and would be a good starting point to realize the advantage of quantum processors over classical computers in solving quantum chemistry problems. Currently, one of the largest quantum processors available on IBM Quantum is the 65-qubit machine \texttt{ibmq\_manhattan}. However, its quantum volume \cite{moll_quantum_2018,cross_validating_2019} is just 32, so a lot of effort still needs to be made to increase the reliable circuit depth, i.e., the number of gates that can be successively and reliably performed in a quantum circuit on the quantum machine.   
In most previous VQE works, the basis sets used are minimal and insufficient to achieve results with the desired accuracy compared with experimental data. 
With QEE, one may perform chemical simulations with larger basis sets but only require a lower amount of qubit resources, which permits a greater accuracy for quantum chemistry calculation in the NISQ era.

Compared to previous compact encoding studies \cite{moll_optimizing_2016, bravyi_tapering_2017, kirby_second-quantized_2021, di_matteo_improving_2021}, our work has a generalized scheme and is practical for NISQ devices, where such qubit reduction can be applied to fermionic systems of any size, and the encoding of two-electron terms is also included. However, the number of Hamiltonian terms
that has to be preprocessed by classical computing seems to be a bottleneck  
in both our work and some compact encoding studies \cite{moll_optimizing_2016, bravyi_tapering_2017, di_matteo_improving_2021, steudtner_fermion--qubit_2018}.
By considering all the possible excitations,
the total number of Pauli operator terms before combining like terms
of our QEE scheme that should be processed \revise{has an upper bound of $\mathcal O(\frac{N^{2m+1}}{(m-1)!\, m!})$} as shown in Appendix \ref{appendix:timecomplexity} due to the decompositions of all entry-operators. \revise{In most cases, the numbers of Hamiltonian terms will be more than those of the JW encoding scheme, which scale as $\mathcal O(N^4)$.}
Nevertheless, Table \ref{tab:JW_QEE_Terms_Comparison} provides several use cases for the QEE scheme, where there are significant reductions of qubit counts and modest numbers of Hamiltonian terms (sometimes even fewer terms than the systems using the JW encoding scheme).


On the other hand, the advantage of 
significant reduction of the qubit number by our QEE scheme for large molecules, which substantially alleviates the vanishing gradient and long ansatz circuit problems, will be useful in the NISQ era if there is a better classical preprocessing method.
Besides, in our proposed QEE scheme, one has the degree of freedom to map certain fermionic states to qubit states in an arbitrary order or basis. This corresponds to choosing suitable basis orders since the decomposition of all entry-operators within a qubit-number-reduced Hamiltonian is equivalent to the tensor product decomposition of that Hamiltonian using Pauli operators.
In this work, we order fermionic states in an ascending manner to respect hardware SPAM errors. 
Some suitably chosen mappings between fermionic states and qubit states may enable us to systematically decompose the encoded excitation operators without expanding all entry-operators. 
In other words, qubit Hamiltonian with different mappings corresponds to different similar matrices so there may exist specific arrangements of the basis to construct qubit Hamiltonian efficiently.
Hence further analysis should be done on other ordering methods to possibly reduce classical preprocessing time or the number of terms in the qubit Hamiltonian.

\section*{ACKNOWLEDGEMENTS}
The authors thank Alice Hu, Yu-Cheng Chen, Peng-Jen Chen, Jyh-Pin Chou, Shih-Kai Chou, and Shou-Yen Hsiao for great suggestions and discussions.
We also thank IBM Quantum Hub at NTU for providing computational resources and accesses for conducting the real quantum device experiments.
H.-C.C. is supported by the Young Scholar Fellowship (Einstein Program) of the Ministry of Science and Technology, Taiwan (R.O.C.) under Grants No.~MOST 109-2636-E-002-001, No.~MOST 110-2636-E-002-009 and No. MOST 111-2119-M-001-004, by the Yushan Young Scholar Program of the Ministry of Education, Taiwan (R.O.C.) under Grants No.~NTU-109V0904 and NTU-110V0904, and by the research project ``Pioneering Research in Forefront Quantum Computing, Learning and Engineering'' of National Taiwan University under Grant No.~NTU-CC-111L894605.
H.-S.G. acknowledges support from the the Ministry of Science and Technology,
Taiwan under Grants No.~MOST 109-2112-M-002-023-MY3, 
No.~MOST 109-2627-M-002-003, No.~MOST 110-2627-M-002-002,
No.~MOST 107-2627-E-002-001-MY3, No.~MOST 111-2119-M-002-006-MY3, 
No.~MOST 111-2119-M-002-007 and No.~MOST 110-2622-8-002-014,
from the US Air Force Office of Scientific Research under
award number FA2386-20-1-4033,
and from the
National Taiwan University under Grant
No.~NTU-CC-111L894604.
H.-C.C. and H.-S.G. are grateful to the support from the National Center for Theoretical Sciences, Physics Division, Taiwan.

\appendix
\section{6-31G $\ch{H2}$ QEE Qubit Hamiltonian} \label{appendix:H2_0.745_QEE_Hamiltonian}
The QEE qubit Hamiltonian for an $\ch{H2}$ molecule in the 6-31G basis set at an interatomic distance of 0.745 Å is
\begin{align} \label{eq: H2-631G-Ham}
\begin{split}
  H_{\text{q}} = & -0.363395 \cdot IIII -0.260044 \cdot IZII \\
    & -0.482367 \cdot ZIII -0.007374 \cdot ZZII \\
    & +0.029427 \cdot XIII -0.061555 \cdot XZII \\
    & -0.260044 \cdot IIIZ -0.482367 \cdot IIZI \\
    & -0.007374 \cdot IIZZ +0.029427 \cdot IIXI \\
    & -0.061555 \cdot IIXZ +0.007946 \cdot IZIZ \\
    & -0.001401 \cdot IZZI +0.004264 \cdot IZZZ \\
    & -0.001401 \cdot ZIIZ +0.010898 \cdot ZIZI \\
    & -0.011880 \cdot ZIZZ +0.004264 \cdot ZZIZ \\
    & -0.011880 \cdot ZZZI +0.025182 \cdot ZZZZ \\
    & +0.001979 \cdot XIIZ -0.004488 \cdot XIZI \\
    & +0.005020 \cdot XIZZ +0.006781 \cdot XZIZ \\
    & -0.021515 \cdot XZZI +0.044350 \cdot XZZZ \\
    & +0.010276 \cdot IXIX -0.011928 \cdot IXZX \\
    & -0.011928 \cdot ZXIX +0.094119 \cdot ZXZX \\
    & +0.005441 \cdot XXIX -0.054641 \cdot XXZX \\
    & -0.016451 \cdot YYIX +0.046704 \cdot YYZX \\
    & +0.001979 \cdot IZXI +0.006781 \cdot IZXZ \\
    & -0.004488 \cdot ZIXI -0.021515 \cdot ZIXZ \\
    & +0.005020 \cdot ZZXI +0.044350 \cdot ZZXZ \\
    & +0.007491 \cdot XIXI +0.010322 \cdot XIXZ \\
    & +0.010322 \cdot XZXI +0.080979 \cdot XZXZ \\
    & +0.005441 \cdot IXXX -0.016451 \cdot IXYY \\
    & -0.054641 \cdot ZXXX +0.046704 \cdot ZXYY \\
    & +0.032133 \cdot XXXX -0.025336 \cdot XXYY \\
    & -0.025336 \cdot YYXX +0.054367 \cdot YYYY.
\end{split}
\end{align}
Note that for brevity, we have ignored the qubit indices of the Pauli terms in Eq. (\ref{eq: H2-631G-Ham}). All terms are labeled with descending indices, e.g.~ $XYZI=X_3Y_2Z_1I_0$.

\section{Derivation of Scaling of $\mathbf{Q}$} \label{appendix:Q_scaling}
As previously mentioned, 
\revise{qubit count}
$Q = \left\lceil
\log_2{N\choose m}
\right\rceil$ \revise{for QEE.} The binomial coefficient can be simplified as
\begin{equation*}
    {N\choose m} 
    = \frac{N\cdot (N-1) \cdots (N-m+1)}{m!} \revise{\ <\ } \frac{N^m}{m!},
\end{equation*}
\revise{so}
\begin{equation*}
    \log_2{N\choose m} \revise{\ <\ } m\log_2(N) - \log_2(m!) = \mathcal O(m\log_2(N)),
\end{equation*}
which is the \revise{upper bound} of $Q$ for the case of $m \revise{\ \leq\frac{N}{2}}$. Similarly, since ${N\choose m} = {N\choose N-m}$,  \revise{ for the case of $m \ >\frac{N}{2}$,
\begin{equation*}
    {N\choose m} 
    = \frac{N\cdot (N-1) \cdots (m+1)}{(N-m)!} \ <\  \frac{N^{N-m}}{(N-m)!},
\end{equation*}
so $Q = \left\lceil
\log_2{N\choose m}
\right\rceil \ <\  \mathcal O\left((N-m)\log_2N\right)$.}


\section{Device Calibration Data} \label{appendix:calibration}

Table \ref{tab:santiago_single} and Table \ref{tab:santiago_CNOT} show the calibration data of a IBM quantum machine \texttt{ibmq\_santiago}. These calibration data were used to construct the noise model used in this work.

\begin{table}[hbt!]
\centering
\caption{The single-qubit calibration data of \texttt{ibmq\_santiago} on March 5, 2021.}
\begin{ruledtabular}
\begin{tabular}{ccccc}
\textrm{Qubit}&
\textrm{Gate error}&
\textrm{Readout error}&
\textrm{$P(0\mid1)$\footnote{ {The probability of measuring the $\ket{0}$ qubit state given that it was prepared in the $\ket{1}$ qubit state.}}}&
\textrm{$P(1\mid0)$\footnote{ {The probability of measuring the $\ket{1}$ qubit state given that it was prepared in the $\ket{0}$ qubit state.}}}\\
\colrule
0     & 0.0228\%                & 1.45\%        & 2.04\% & 0.86\%  \\
1     & 0.0183\%                & 1.34\%        & 1.42\% & 1.26\%  \\
2     & 0.0217\%                & 8.00\%        & 1.66\% & 14.34\% \\
3     & 0.0262\%                & 3.36\%        & 4.20\% & 2.52\%  \\
4     & 0.0174\%                & 0.89\%        & 1.48\% & 0.30\%  \\
\end{tabular}
\end{ruledtabular}

\label{tab:santiago_single}
\end{table}

\begin{table}[hbt!]
\centering
\caption{The CNOT gate calibration data of \texttt{ibmq\_santiago} on March 5, 2021.}
\begin{ruledtabular}
\begin{tabular}{ccc}
Coupling   Pair & Gate error & Gate length (ns) \\
\hline
{[}0, 1{]}      & 0.573\%         & 526.22                \\
{[}1, 0{]}      & 0.573\%         & 561.78                \\
{[}1, 2{]}      & 0.686\%         & 604.44                \\
{[}2, 1{]}      & 0.686\%         & 568.89                \\
{[}2, 3{]}      & 0.670\%         & 376.89                \\
{[}3, 2{]}      & 0.670\%         & 412.44                \\
{[}3, 4{]}      & 0.636\%         & 376.89                \\
{[}4, 3{]}      & 0.636\%         & 341.33                \\
\end{tabular}
\end{ruledtabular}

\label{tab:santiago_CNOT}
\end{table}

\section{Propagation of Uncertainties} \label{appendix:uncertainties}
The error bars for the extrapolated energies at different interatomic distances in Figs. \ref{fig:H2_VQE_10_Error}, \ref{fig:LiH_VQE_10_Error} and \ref{fig:No_VQE} are
calculated by assessing the propagation of uncertainties from the standard deviations $\sigma_i$'s of the experiments for different CNOT gate counts $x_i$'s and their linear regression residues. The uncertainty $\sigma$ of the y-intercept (in a energy versus CNOT gate count plot) can be written as
\begin{equation*}
    \sigma = \sqrt{\frac{1}{\Delta}\sum_{i}\frac{x_i^2}{\sigma_i^2}},
\end{equation*}
where $\Delta=(\sum_{i}\frac{1}{\sigma_i^2})(\sum_{i}\frac{x_i^2}{\sigma_i^2})-(\sum_{i}\frac{x_i}{\sigma_i^2})^2$. Note that the 95\% confidence intervals of the distributions are plotted as the error bars in Figs.~\ref{fig:H2_VQE_10_Error}, \ref{fig:LiH_VQE_10_Error} and  \ref{fig:No_VQE} so the lengths of error bars are $2\sigma$.

\section{Complexity of Hamiltonian Encoding with Entry-Operators} \label{appendix:timecomplexity}
To obtain the qubit Hamiltonian $H_\text{q}$ corresponding to the second-quantized electronic Hamiltonian $H_{\text{elec}}$, one has to first calculate qubit counterparts of all possible excitation operators $E_{pq}$. 
Each $E_{pq} = \sum_{k,k'=0}^{|\mathcal F _m|-1} 
c_{k'k}^{pq} \ket{\mathbf f_{k'}}_{\text{f}} \bra{\mathbf f_k}_{\text{f}}$
is a linear combination of various fermionic state transitions. 
The number of transitions with non-zero coefficients in $E_{pq}$ is ${{N-1}\choose {m-1}}$ for $p=q$ and ${{N-2}\choose {m-1}}$ for $p\neq q$. Thus, the total number of transitions needs to be calculated for all $E_{pq}$ is $N\times {{N-1}\choose {m-1}} + N(N-1)\times {{N-2}\choose {m-1}}$. 
If \revise{$m \leq \frac{N}{2}$}, one has ${N\choose m} \revise{\ <\frac{N^m}{m!}}$. 
Then the total number of transitions \revise{has an upper bound of $\mathcal \revise{O(\frac{N^{m+1}}{(m-1)!})}$.} 
For each transition $\ket{\mathbf f_{k'}}_{\text{f}} \bra{\mathbf f_k}_{\text{f}}$, it is mapped to a qubit transition $\ket{\mathbf q_{k'}}_{\text{q}} \bra{\mathbf q_k}_{\text{q}}$ and then factorized into a product of $Q$ entry-operators. 
As an entry-operator is a sum of two Pauli (or identity) operators, expanding the product gives $2^Q$ terms of Pauli operators. 
To derive all excitation operators in terms of Pauli operator strings with brute force, $\mathcal O(2^Q N^{m+1})$ Pauli operator terms have to be calculated classically. 
In the case where \revise{$m \leq \frac{N}{2}$}, $Q$ \revise{has an upper bound of $\log_2(\frac{N^m}{m!})$,} so the total number of Pauli operator terms before combining like terms \revise{has an upper bound of $\mathcal O(\frac{N^{2m+1}}{(m-1)! \, m!})$.  Similar results can be obtained for $m\ >\frac{N}{2}$ using the relation of ${N\choose m} = {N\choose N-m}$}, \revise{and in this case the total number of Pauli operator terms
  has an upper bound of $\mathcal O(\frac{N^{2(N-m)+1}}{(N-m-1)! \, (N-m)!})$.
In most cases, the number of Pauli operator terms that need to be processed for QEE is larger than $\mathcal O(N^{4})$ for JW encoding. }


\bibliography{new_bib_r1.bib}

\end{document}